\documentclass[preprint2]{aastex}
\usepackage{epsfig}
\usepackage{natbib}
\usepackage{amsmath}

\begin{document}
\newcommand{\kms}{km~s$^{-1}$}
\newcommand{\Msun}{M_{\odot}}
\newcommand{\Lsun}{L_{\odot}}
\newcommand{\ML}{M_{\odot}/L_{\odot}}
\newcommand{\etal}{{et al.}\ }
\newcommand{\hhh}{h_{100}}
\newcommand{\hsq}{h_{100}^{-2}}
\newcommand{\tn}{\tablenotemark}
\newcommand{\mdot}{\dot{M}}
\newcommand{\p}{^\prime}
\newcommand{\kmsMpc}{km~s$^{-1}$~Mpc$^{-1}$}

\title{Galaxy Groups: A 2MASS Catalog}

\author{R. Brent Tully,}
\affil{Institute for Astronomy, University of Hawaii, 2680 Woodlawn Drive,
 Honolulu, HI 96822, USA}

\begin{abstract}
\noindent
A galaxy group catalog is built from the sample of the 2MASS Redshift Survey almost complete to $K_s=11.75$ over 91\% of the sky.  Constraints in the construction of the groups were provided by scaling relations determined by close examination of well defined groups with masses between $10^{11}$ and $10^{15}~\Msun$.  Group masses inferred from $K_s$ luminosities are statistically in agreement with masses calculated from application of the virial theorem.  While groups have been identified over the full redshift range of the sample, the properties of the nearest and farthest groups are uncertain and subsequent analysis has only considered groups with velocities between 3,000 and 10,000 \kms.   The 24,044 galaxies in this range are identified with 13,607 entities, 3,461 of them with two or more members.  A group mass function is constructed.  The Sheth$-$Tormen formalism provides a good fit to the shape of the mass function for group masses above $6h^{-1} \times 10^{12}~\Msun$ but the count normalization is poor.  Summing all the mass associated with the galaxy groups between 3,000 and 10,000 \kms\ gives a density of collapsed matter as a fraction of the critical density of $\Omega_{collapsed} = 0.16$.

\bigskip
\end{abstract}

\section{Introduction}

The definition of galaxy groups is fraught with ambiguity.  Practitioners of simulated N-body universes have a relatively easy time, given precise knowledge of the three-dimensional positions and velocities of large numbers of test particles.  With observations of the real universe only the two projected dimensions of galaxy positions are known with high accuracy.  The third dimension, distance, is usefully measured only for the nearest systems.  Radial velocities are only crude discriminants, strongly modified by non-linear gravitational effects.  Of proper motions we generally know nothing.  Then there are the problems of small numbers and lost information.  Most galaxies live in groups that contain only a few members substantial enough to be recorded at a significant distance.  As distance increases the recorded fraction of group members drops.

The challenge to a cataloger is to capture the identities of groups consistently over wide ranges of both group mass and distance.  Commonly, group constructions are built with a friends-of-friends algorithm \citep{1982ApJ...257..423H}.  The information is pairwise separations in projection and radial velocity.  The spatial separation scale chosen for the group construction is clearly dependent on distance since the density of recovered candidates falls with distance.  Simultaneously, physically appropriate separation scales vary with group mass.  Similarly, but even more sensitively, group velocity differentials depend on mass.  It is possible to incorporate the additional information of galaxy luminosities into the linkages.  A recipe relating light to mass can be introduced.

A catalog can be built under a set of assumptions about the coupling of separations in space and velocity across large ranges in mass and distance.  Different assumptions may achieve inclusion of true group members and rejection of interlopers with varying degrees of success.  Ultimately at issue is to what degree the properties of the cataloged groups are a correct representation of nature as opposed to a biased representation imposed by the assumptions within the group-finding algorithm.  There have been many attempts to build group catalogs since an early qualitative effort by \citet{1975gaun.book..557D}.   Many involve variations of friends-of-friends linkages \citep{1982ApJ...257..423H, 1983ApJS...52...61G, 1989ApJS...69..809M, 2000ApJ...543..178G, 2002AJ....123.2976R, 2002MNRAS.335..216M, 2004MNRAS.348..866E, 2007ApJ...655..790C, 2011MNRAS.416.2840L, 2014A&A...566A...1T}.  Others construct trees and/or use luminosities as a criterion for connectivity \citep{1978A&A....63..401M,  1984TarOT..73....1V, 1987ApJ...321..280T,  1987MNRAS.225..505N, 1993ApJS...85....1N,  2005MNRAS.356.1293Y, 2006MNRAS.366....2W, 2011MNRAS.412.2498M}.  This new catalog is built with a methodology that most closely resembles that by \citet{2005MNRAS.356.1293Y}.

The present project is motivated by the proposition that we have a basic understanding of what constitutes a group so the search parameters should conform to this knowledge.  Clearly there is a risk that the constraints will prejudice the outcome.  However it can be argued that all extant group catalogs have been prejudiced by their choices of constraints.  Indeed in all extant group catalogs it is not hard to find proposed groups that are questionable.  A test of the current catalog will be whether it is a better survivor of this critique.

A companion paper \citep{2015AJ....149...54T}, hereafter T15, presents the rationale for the group-finding algorithm that will be used.  In that study, a modest number of very well studied groups are given attention.  Those groups range from the Coma Cluster at the high mass end, at $\sim 10^{15} \Msun$, to associations of dwarf galaxies with masses of a few times $10^{11} \Msun$.  Some of the groups are overwhelmingly dominated by early-type HI gas-poor systems and others are dominated by late-type gas-rich systems.  In these well studied cases there are signatures of the extent of the (quasi) virialized halos and infall regions.  It was possible to quantify scaling relations between dimensions, velocity dispersions, and masses.  It is found that mass-to-light ratios for groups increase with group mass \citep{2005ApJ...618..214T, 2009ApJ...695..900Y}.  These group characteristics seem well enough established that they ought to be used as constraints in the construction of a group catalog. 

The galaxies that will be considered for the current group catalog are drawn from the 2MASS Redshift Survey (2MRS) sample that, aside from a narrow low Galactic latitude exclusion zone, is almost complete to $K_s = 11.75$ \citep{2012ApJS..199...26H}.  \citet{2007ApJ...655..790C} constructed a group catalog with a 2MRS sample limited at $K_s = 11.25$ and \citet{2011MNRAS.416.2840L} built a 2MASS catalog with flux limits that differ across regions of the sky.

The methodology that will be discussed involves a translation from luminosities to masses.  Since the input catalog is flux limited, an adjustment is required to account for lost light with distance.  
All galaxies in the 2MRS 11.75 sample are given group assignments but the physical nature of the groups at expansion velocities above 10,000 \kms\ are suspect.
The lost light adjustment flairs to large values at large distances, begging the imposition of an upper limit to the practical catalog at 10,000 \kms\ in the CMB frame 

There is also a practical low velocity limit of 3,000 \kms\ (CMB frame).  Relative group parameters are determined from redshifts (assuming H$_0 = 100 h$ \kmsMpc) and peculiar velocities add scatter to parameters at low observed velocities.  Also, there are interesting options to a 2MRS sample locally.  2MASS samples misses low surface brightness galaxies.  Such galaxies do not make an important contribution beyond 3,000 \kms\ but are dominant in a nearby sample \citep{2013AJ....145..101K}.  At present, perhaps the best group catalog for the volume within 3,000 \kms\ is the one by \citet{2011MNRAS.412.2498M}.  The ambition here is to provide the best group catalog for the volume between 3,000 and 10,000 \kms.  All galaxies in 2MRS 11.75 are given an assignment reflecting their environment, be that to a group of two or more members or as a ``group'' of one.

The outline of this paper is as follows.  In the next section there will be a review of what can be claimed to be understood about the properties of galaxy groups, drawing heavily on the companion paper T15.  The third section will discuss the 2MRS 11.75 base catalog.  There will be an evaluation of completeness issues, then a formulation of a group-finding algorithm consonant with the ideas presented in section 2.  Section four hosts the catalog.  In section five there is a discussion of the statistics of group properties.  A group mass function is constructed in section six.

\section{Properties of Known Galaxy Groups} 

Dimension parameters that are accessible to N-body simulators, like the overdensity scale $r_{200}$, a radius at 200 times the density required to close the universe with matter, or the dynamical gravitational radius $r_g$
\begin{equation}
r_g = {\Sigma_{i>j} m_im_j\over\Sigma_{i>j} m_im_j / r_{ij}}
\label{eq:rg}
\end{equation}
are not very useful to observers.  How does one define a density surface about groups with only a few members?  Likewise, the projected gravitational radius is unstable if statistics are poor and, in addition, requires a postulation of group membership to be applied.

It was argued in the companion paper T15 that a group manifests a scale that is accessible to observations: the projected radius of second turnaround, $R_{2t}$ (the nomenclature of T15 is followed, with projected dimensions in upper case $R$ and those in 3D in lower case $r$).  With radial infall in a spherical system, galaxies initially plunge through the group and then continue outward until they stall at the second turnaround radius, $r_{2t}$.  Simplest models anticipate a cusp at $r_{2t}$ \citep{1985ApJS...58...39B,1989RvMP...61..185S}.  With N-body collapse models the cusp is damped but a density discontinuity feature remains \citep{2009MNRAS.400.2174V}.

The discontinuity associated with $R_{2t}$ can be seen in judiciously selected groups.  Here ``judicious'' means the chosen groups have been studied individually in detail, both with imaging and spectroscopy, to depths sufficient to generate large samples.  Efforts were made to minimize interloper problems.  Details are given in T15.

A second parameter readily measured for these well studied groups is the radial component of the velocity dispersion $\sigma_p$ for galaxies within the radius $R_{2t}$.  In T15 it is demonstrated that there is a linear relationship between the two measured parameters \footnote{The scaling relations cited in T15 are derived with distance measures consistent with H$_0=75$~\kmsMpc.  In this paper, distances are converted from velocity units assuming H$_0=100$~\kmsMpc\ and the scaling relations are modified accordingly.  Following the standard convention, $h={\rm H}_0/100$.}
\begin{equation}
\sigma_p / R_{2t} = 491\pm11 h~{\rm km ~s^{-1} ~Mpc^{-1}}
\end{equation}
established from consideration of 13 groups ranging from the M31 halo to the Coma Cluster.

Given a census of group members within $R_{2t}$, the projected gravitational radius can be calculated.  Mass weighting, as explicit in Eq.~\ref{eq:rg}, is not a good idea because a mass-weighted solution is dominated by a small number of objects and is unstable.  The information from dwarf test particles is ignored.  Instead, the unweighted formulation is used
\begin{equation}
R_g = {N^2\over\Sigma_{i>j} 1/R_{ij}}
\end{equation}
where there are N group candidates and pair projections are counted only once.  With this information, the group virial mass, $M_{\rm v}$, can be calculated
\begin{equation}
M_{\rm v} = \sigma_{3D}^2 r_g / G = (\alpha\pi/2G) \sigma_p^2 R_g
\end{equation}
where statistically $r_g = (\pi/2)R_g$ and $\sigma_{3D} = \sqrt{\alpha} \sigma_p$.  The parameter $\alpha$ depends on the nature of the orbits and in T15 there is justification for the choice $\alpha = 2.5$.

With a dynamical measure of $M_{\rm v}$ in hand (or $M_{12} = M_{\rm v} / 10^{12} \Msun$), T15 demonstrated the correlation
\begin{equation}
R_{2t} = 0.178\pm0.005 h^{-2/3} M_{12}^{1/3} ~{\rm Mpc}
\label{eq:r2t}
\end{equation}
with 8 well studied groups ranging in mass from $2 \times 10^{12} \Msun$ to $2 \times 10^{15} \Msun$. An alternate, not independent, formulation is
\begin{equation}
M_{\rm v} = 1.5 h^{-1} \times 10^6 \sigma_p^3 ~.
\label{eq:sigp}
\end{equation}

How can this information be used to define groups from a redshift catalog?  Given knowledge of group mass then the projected group scale $R_{2t}$ and velocity dispersion $\sigma_p$ would be statistically known.  Initially there is no knowledge of group mass but there is information from a proxy: $K$ band luminosity.  It is next necessary to evaluate the relation $M_{\rm v} / L_K$ for groups over a range of masses.

As a starting point, there is a known dependence of mass-to-light ratio on group mass at optical bands.  \citet{2005ApJ...618..214T} found the dependence in blue light $L_B \propto M_{\rm v}^{0.7}$ at masses above $10^{12} \Msun$.  Group luminosities drop off rapidly below $10^{12} \Msun$ but this is a regime that will not be of concern here.  \citet{2002ApJ...569..101M} and \citet{2003MNRAS.345..923V} reached similar conclusions indirectly from the need to reconcile the galaxy luminosity function and the halo mass function.

At $K_s$ band the Coma Cluster provides a robust mass-to-light calibration at high mass.  T15 gave attention to a deeper survey than 2MRS 11.75, the 2MASS Extended Source Catalog  \citep{2000AJ....119.2498J} with 367 group redshifts.    At a distance given by the group mean CMB frame redshift of 7331 \kms\ and H$_0 = 100 h$ \kmsMpc\ (as will be assumed consistently in this discussion) the Coma Cluster projected second turn-around radius is $R_{2t} = 2.2 h^{-1}$ Mpc, the projected velocity dispersion is $971\pm51$ \kms, the group virial mass is $M_{\rm v} = 1.7 h^{-1} \times 10^{15} \Msun$, and the $K_s$ luminosity is $L_{K_s}=1.4 h^{-2} \times 10^{13} \Lsun$, whence $M_{\rm v}/L_{K_s} = 120 h \Msun/\Lsun$.  Here, $L_{K_s}$ has received adjustments for incompletion as will be discussed in the next section.

At the other extreme of the mass range of interest, there is good information from the numerical action modeling of galaxy orbits in the vicinity of nearby groups \citep{2013MNRAS.436.2096S}.  A characteristic value for the dominant spirals in small groups is $M/L_{K_s} = 40 h \Msun/\Lsun$.  Elliptical galaxies have 25\% more mass per unit $K_s$ light.

This difference in $M/L$ over the interval $10^{12} - 10^{15} \Msun$ is consistent with the dependence found by \citet{2011MNRAS.412.2498M} of $M_p \propto L_{K_s}^{1.15}$ where $M_p$ is the projected mass \citep{1985ApJ...298....8H}.  These consistent constraints motivates the following formula for the translation from observed luminosity to expected group mass $M_{\rm v}^{exp}$
\begin{equation}
M_{\rm v}^{exp} = 43 \times 10^{10} (L_{10})^{1.15} \Msun 
\label{eq:ml}
\end{equation}
where $L_{10}$ is the $K_s$ luminosity in units of $10^{10} \Lsun$ and assuming H$_0=100$~\kmsMpc.  This equation leads to $M_{\rm v}/L_{K_s} = 43~ \Msun/\Lsun$ at $L_{K_s} = 10^{10} \Lsun$ and $121 ~\Msun/\Lsun$ at $L_{K_s} = 10^{13} \Lsun$. Below and above these luminosity limits $M_{\rm v}/L_{K_s}$ is held constant at 43 and 121 $\Msun/\Lsun$ respectively.  The uncertainty in this transformation from luminosity to mass is estimated at 20\% (68\% probability) from comparisons between virial and luminosity masses discussed in Section~\ref{sec:gpprop}.

\section{The 2MRS 11.75 Catalog}

The 2MASS Redshift Survey (2MRS) with redshifts available for 98\% of galaxies from the Two Micron All Sky Survey \citep{2000AJ....119.2498J, 2003AJ....125..525J} brighter than $K_s = 11.75$ outside a Milky Way exclusion zone of 9\% of the sky is the last publication and tribute to John Huchra \citep{2012ApJS..199...26H}.  The redshift catalog used here contains 43,065 entries.  In the following, 2MRS total magnitudes corrected for extinction are used, resulting in a completion limit that is slightly boosted to $K_s=11.5$.

\subsection{The 2MRS Luminosity Function}

As with any flux limited sample, intrinsically faint galaxies are lost with increasing distance, so a first order of business is to determine a selection function that describes this loss of information.  Others have considered this problem with related data sets.  Here, a blind formulation is derived and then compared with the literature.

It is to be appreciated that the 2MASS catalog misses low surface brightness galaxies.  The 2MASS exposures were short, observing with modest sized telescopes, against high and variable auroral sky emission in the $K_s$ band.  The photometric integrity of the survey is exceptional but Figure 1 provides a demonstration, in comparison with a deeper survey of a small region \citep{1996AJ....112.2471T}, of the loss of low surface brightness flux.  Very low surface brightness galaxies are lost entirely from the 2MASS catalog.

\begin{figure}[htbp]
\begin{center}
\includegraphics[scale=0.4]{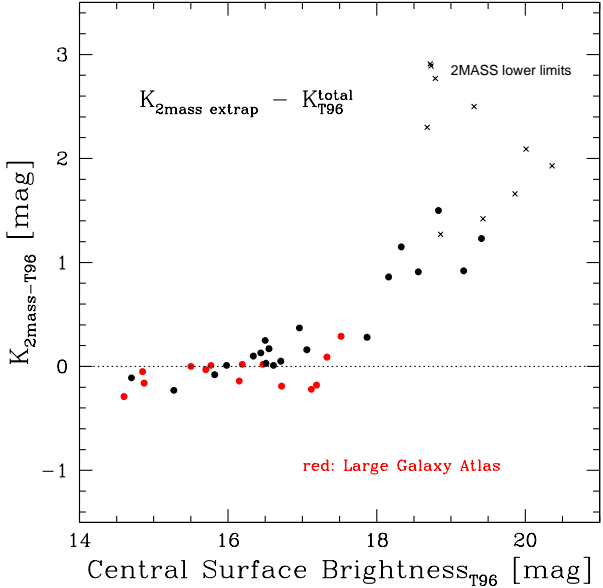}
\caption{Demonstration of lost flux from low surface brightness galaxies in the 2MASS catalog for galaxies in the nearby Ursa Major Cluster.  On the ordinate, the magnitude difference is shown between the 2MASS $K_s$ magnitude and the magnitude measured in the more sensitive observations by \citet{1996AJ....112.2471T}.  The exponential disk central surface brightness of a galaxy is given on the abscissa.  Symbols in red represent galaxies that appear in the Large Galaxy Atlas \citep{2003AJ....125..525J} which is more rigorous than the Extended Source Catalog \citep{2000AJ....119.2498J}. The crosses identify lower limits to the magnitude differentials in cases of galaxies excluded from the 2MASS catalog. The 2MASS survey looses flux with a strong dependence on surface brightness.}
\label{lsb}
\end{center}
\end{figure}

Although Figure~\ref{lsb} looks scary, the loss of low surface brightness galaxies is not a substantial problem for the present study because a lower redshift limit is set for the `practical' group catalog that is being assembled (the regime 3,000$-$10,000~\kms).  Low surface brightness galaxies are numerically dominant in a local volume limited sample to faint magnitudes but they make a small contribution to integrated flux.  Here, a 2MASS $K_s$ band luminosity function is constructed to a limiting magnitude of $M_{K_s} = -19 + 5{\rm log} h$.  The contribution from galaxies fainter than this limit would augment the total luminosity budget by 2-5\% depending on the steepness of the faint-end luminosity function.  Of slightly greater concern is the missing flux from low surface brightness galaxies above the magnitude cutoff.  As a rough estimate, as much as 20\% of the light in stars in galaxies might be missing from our inventory.  

In defining the luminosity function it is necessary to contend simultaneously for the loss of faint galaxies with distance and the uneven contributions with distance due to clumping in the distribution of galaxies.  In the regime where peculiar velocities are only a small perturbation on expansion velocities, it is possible to deal with the clumping problem by fitting the luminosity function complete to a limit defined by the inner redshift boundary of the shell.  A parameterized function can be fit to the observed luminosity distribution to the completeness limit for a sequence of shells over a domain relevant to the input catalog.  If the parameters of the fits are stable over successive shells then it can be considered that the luminosity function is reasonably defined.

A Schechter function \citep{1976ApJ...203..297S} is fit following standard conventions.  The function for absolute magnitude $M_{K_s}$ is of the form $N(M_{K_s}) =$
\begin{equation}
N_K {\rm exp} [ -10^{-0.4(M_{K_s}-M_K^{\star})}] 10^{-0.4(\alpha_K+1)(M_{K_s}-M_K^{\star})}
\end{equation}
and has three fitting parameters: a faint end power law slope parameter $\alpha_K$, a bright end exponential cutoff parameter $M_K^{\star}$, and a sample normalization parameter $N_K$.  Only the first two parameters are important for the present discussion.

It has to be entertained that galaxy luminosity functions may vary with environment.  Detailed studies suggest that the variations are, in fact, quite modest \citep{2011EAS....48..281T}.  Variations in $\alpha$ at $r$ band range from $-1.35$ for dense, gas-poor environments to $-1.2$ for less evolved gas-rich environments.  There is a coupling in the parameter fits: a brighter choice of $M^{\star}$ leads to a more negative choice of $\alpha$.  There is tentative evidence at optical bands that there is a weak environmental dependence on luminosity functions but the more robust conclusion is that any dependence is at most small.

Before giving attention to the full 2MRS catalog in redshift shells, three nearby clusters are considered that completely sample to the $M_{K_s} = -19.0$ limit: the Virgo, Fornax, and Ursa Major clusters (here, luminosities have been transformed to be consistent with H$_0=100$).  With the Virgo Cluster alone, a sample of 150 galaxies is fit with the $M_K^{\star}$ and $\alpha_K$ parameters $-23.7$ and $-0.97$.  Extending the sample to include 234 galaxies in the three clusters gives $-23.4$ and $-0.83$.  In comparing these numbers, recall the cautionary note above regarding parameter coupling.

Turning to the full sample, consideration is given to fits in 15 separate shells at half magnitude increments of $M_K^{lim}$, describing the faintest galaxies accessed in shells over the range $-19 \leq M_K^{lim} \leq -26$.  The corresponding inner velocity limits range from 1259 \kms\ to 30,000 \kms.  (With completion to $K_s = 11.5$ then there is a complete representation of the luminosity function to $M_{K_s} = -19 +5{\rm log} h$ for the volume within expansion velocities of 1259 \kms.)
The nearest shells contain $\sim 500$ galaxies.  The most populated shell at 10,000 \kms\ contains almost 8,000 galaxies.

The best constraints on the faint end slope $\alpha_K$ come from fits to the nearest shells.  However in these cases the bright end is poorly populated, creating uncertainty in $M_K^{\star}$.  The bright end exponential cutoff parameter $M_K^{\star}$ is most confidently constrained in the mid range of shells which are well populated.  In the five most distant shells beyond 10,000 \kms, $M_K^{lim} < M_K^{\star}$, so $\alpha_K$ is effectively unconstrained and the fits are affected by rare high luminosity systems.

Giving attention to the nearest shells there is the hint that the luminosity function is turning over ($\alpha_K < -1$) but the evidence is not convincing.  This hint was already seen with the fits to the three nearby clusters.  However the details of the faint end slope are not important here.  At the near limit of 3,000 \kms\ of the `practical' group catalog that will be constructed, $M_K^{lim} = -20.9$.  For the ensuing fits $\alpha_K = -1.0$ is assumed and the only free parameter is $M_K^{\star}$.  From the discussion earlier in this section it is to be appreciated that the faint end slope measurement for the 2MASS catalog is strongly biased shallow by the loss of low surface brightness galaxies.  

\begin{figure}[htbp]
\begin{center}
\includegraphics[scale=0.4]{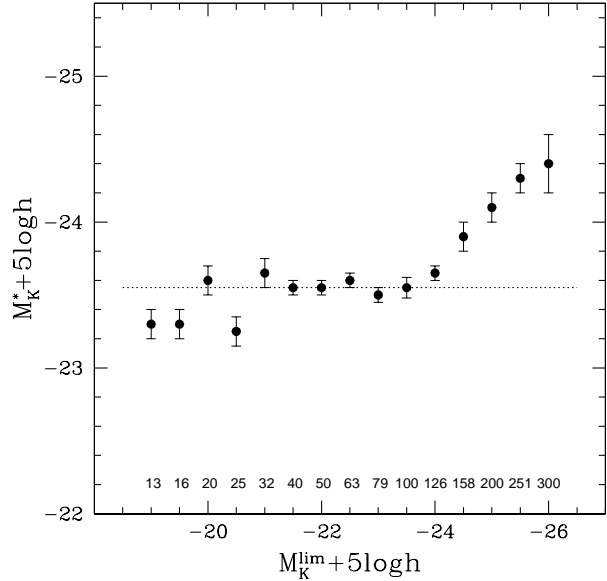}
\caption{Best fits for $M_K^{\star}$ in discrete redshift shells assuming $\alpha_K=-1.0$.  Expansion velocities at the inner edge of shells are given in units of 100 \kms\ along the bottom of the figure.  Shells at velocities above 10,000 \kms\ only sample galaxies brighter than $M_K^{\star}$.}
\label{mlms}
\end{center}
\end{figure}

Figure \ref{mlms} shows the run of fits to $M_K^{\star}$ as a function of the magnitude limit $M_K^{lim}$ in shells, assuming $\alpha_k = -1.0$.  In the four most distant shells, at velocities $> 16,000$ \kms, there is a drift in $M_K^{\star}$ to higher luminosity.  These shells only sample the brightest galaxies in the exponential cutoff tail.  There is a hint here that the Schechter description is inadequate, that there is an excess of extremely luminous galaxies.  Indeed, there is evidence from a more extensive sample derived from a larger volume that there are more bright galaxies than anticipated by the exponential cutoff of the Schechter function and that a better description is provided by a double power law \citep{2009A&A...495...37T}. That possibility is not of concern to the current discussion given an upper expansion velocity cutoff of 10,000 \kms\ to the practical group catalog.

In summary of this section, the luminosity function is adequately fit by the exponential cutoff parameter $M_K^{\star} = -23.55 \pm 0.05 + 5{\rm log} h$ assuming $\alpha_K = -1.0$.  By comparison, \citet{2001MNRAS.326..255C} cite $-23.44 \pm 0.03$ with $\alpha_K = 0.96 \pm 0.05$ from 17,173 galaxies with redshifts from the 2dF survey, \citet{2003ApJS..149..289B} cite $-23.33$ with $\alpha_K = -0.88$ brightward of $-21 + 5{\rm log} h$ and $\alpha_K = -1.33$ faint-ward, from 22,679 galaxies with redshifts from the Sloan survey, and \citet{2007ApJ...655..790C}, referencing Huchra, cite $-23.5$ with $\alpha_K = -1.02$ from the preliminary 2MRS 11.25 sample.  These earlier results are consistent with the present results given the issue of the lost low surface brightness systems and the coupling between the parameters $M_K^{\star}$ and $\alpha_K$.

\subsection{The Selection Function} 

\begin{figure}[htbp]
\begin{center}
\includegraphics[scale=0.4]{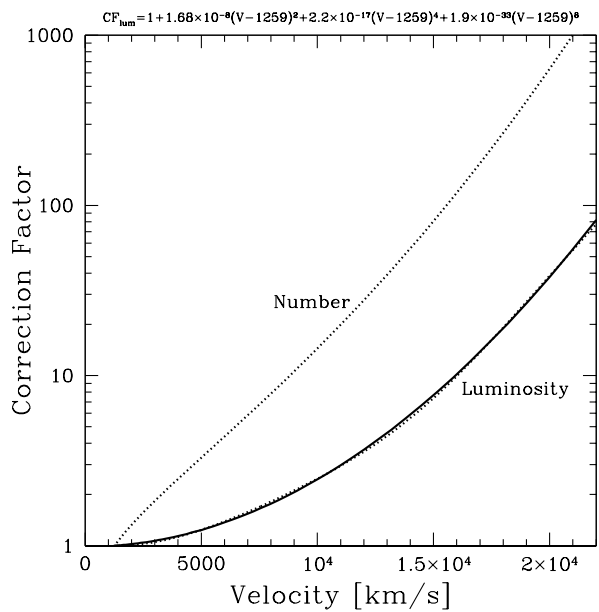}
\caption{Number and luminosity correction factors as a function of velocity in the Local Sheet reference frame.  Eq.~\ref{Eq:cf} for $CF_{lum}$ is described by the solid line, overlying a dotted line that plots the numerically calculated run of this parameter.}
\label{cf}
\end{center}
\end{figure}

Given a luminosity function, the number and flux from missing galaxies can be determined as $M_K^{lim}$ increases in brightness with distance.  Figure~\ref{cf} illustrates this lose with expansion velocity.  Contributions faintward of $M_{K_s} = -19.0 + 5{\rm log} h$ are ignored so there is complete coverage at 1259 \kms.  The correction factor is a multiplier to account for missing information.  The number of missing galaxies explodes quickly but the lose of light increases much less rapidly.  The luminosity correction factor $CF_{lum}$ is described by the polynomial expression
\begin{multline}
CF_{lum} = 1 + 1.68\times10^{-8}(V_{LS}-1259)^2  \\+ 2.2\times10^{-17}(V_{LS}-1259)^4 + 1.9\times10^{-33}(V_{LS}-1259)^8
\label{Eq:cf}
\end{multline}
where velocities $V_{LS}$ are in the Local Sheet frame \citep{2008ApJ...676..184T}, a variation on the more familiar but ambiguous Local Group frame.  By 10,000 \kms, where $M_L^{lim} \simeq M_K^{\star}$, the correction factor for lost light is about a factor two.

A procedure will be described in the next section that requires luminosities for the definition of galaxy groups.  If $CF_{lum} \gtrsim 2$ the composition of groups become suspect with the proposed procedure.  Consequently, although the full 2MRS 11.75 catalog is given attention and groupings are proposed for the ensemble, group compositions are not considered reliable at velocities greater than 10,000 \kms.  At the other extreme, because of issues mentioned earlier associated with peculiar velocities and missing low surface brightness objects, the current catalog is not the best below 3,000 \kms.  The domain of the `practical' catalog is 3,000$-$10,000~\kms.  The properties of groups outside this regime are not considered reliable.

\section{Construction of the Group Catalog}

Observable properties of established galaxy groups were discussed in section 2, including second turnaround dimensions $R_{2t}$ distinguishable in well studied cases, the line-of-sight velocity dispersion $\sigma_p$ in clean cases, and the associated luminosities.  Scaling laws were established that permit the inference of halo properties $-$ mass, velocity dispersion, and radius - from observed luminosities.  Section 3 provided a description of adjustments that have to be made to luminosities to account for lost contributions as a function of distance.

The conceptual outline of the group-finding algorithm is as follows.  (1) Start with the intrinsically most luminous galaxy in the sample after adjustment with the correction factor.  (2) Assume a group mass-to-light ratio appropriate for that intrinsic luminosity using Eq.~\ref{eq:ml} and calculate the halo expectation parameters $R_{2t}$ (Eq.~\ref{eq:r2t}) and $\sigma_p$ (Eq.~\ref{eq:sigp}).  (3) Cycle through the sample to search for galaxies that lie within the $R_{2t}$ radius of the primary system and within $2\sigma_p$ of its velocity.  (4) After this first cycle, sum the luminosities of associated galaxies and determine their luminosity weighted projected centroid and unweighted velocity mean.  (5) Calculate the halo expectation parameters $R_{2t}$ and $\sigma_p$ for this enlarged entity.  Repeat cycles until there are no new links.  (6) Go to the next intrinsically most luminous galaxy among the unlinked cases and repeat procedures $(2)-(5)$.  (7) Continue to successively fainter galaxies until there are no more galaxies to consider.

After the initial construction of tentative groups, it is found that there can be occasional overlaps between close neighbors.  Hence another cycle is initiated.  (8) Beginning with the most populated candidate group, cycle through the other groups looking for overlaps in both $R_{2t}$ projected dimensions and velocity dispersions (the larger of the quadrature addition of the $2\sigma_p$ values and the $3\sigma_p$ value of the larger entity).  (9) Recalculate halo properties for any enlarged candidate group and recycle.  Repeat until no new additions.  (10) Consider next most populous candidate group and successively smaller candidates until the entire catalog has been explored.  The final affiliations of galaxies will be called "nests". 

\begin{figure}[htbp]
\begin{center}
\includegraphics[scale=0.4]{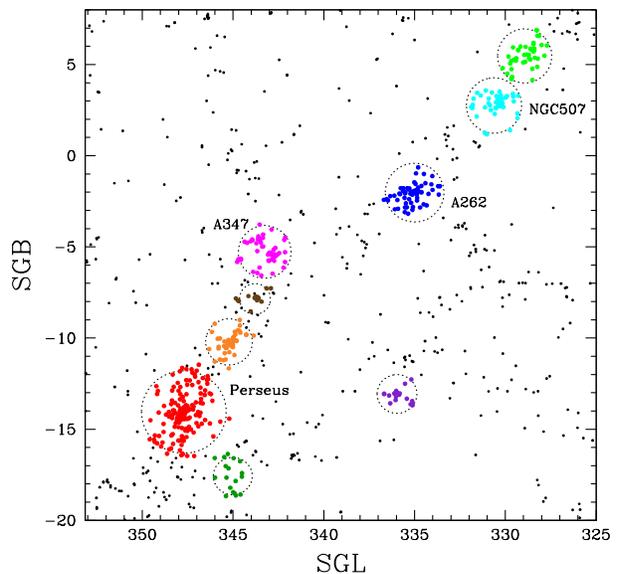}
\caption{The Perseus-Pisces filament.  Major components are given distinct colors.  Dotted circles indicate $R_{2t}$, second turnaround radii for the major halos.  Points in black identify all other 2MRS $K<11.75$ galaxies with $3500<V<6500$ \kms.}
\label{lbpp}
\end{center}
\end{figure}

\begin{figure}[htbp]
\begin{center}
\includegraphics[scale=0.4]{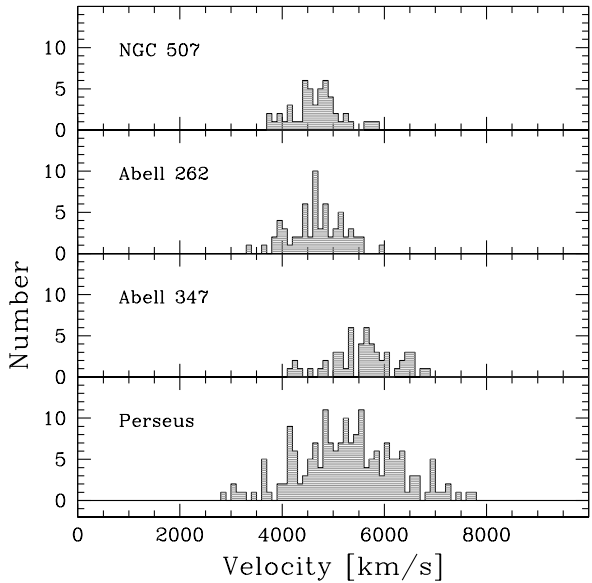}
\caption{Velocity histograms for the four most important components of the Perseus-Pisces filament.}
\label{vpp}
\end{center}
\end{figure}

An illustration of groups found in a particular region is given in Figure~\ref{lbpp}.  The Perseus-Pisces filament that is shown is the densest region of major filaments in the volume that has been explored.  Histograms of the velocities of candidate members in the four largest nests are seen in Figure~\ref{vpp}. 

Among details, one is a cutoff in the galaxies that are considered at an expansion velocity of 24,000 \kms.  The idea is to identify groups in some manner to 20,000 \kms\ with minimal high velocity edge effect.  As mentioned already, though, the luminosity correction factor becomes large at high redshifts, reaching 40 at 20,000 \kms.  The group catalog extends to 20,000 \kms\ but it would be unwise to make much of group characteristics at velocities greater than 10,000 \kms.

\begin{figure}[htbp]
\begin{center}
\includegraphics[scale=0.4]{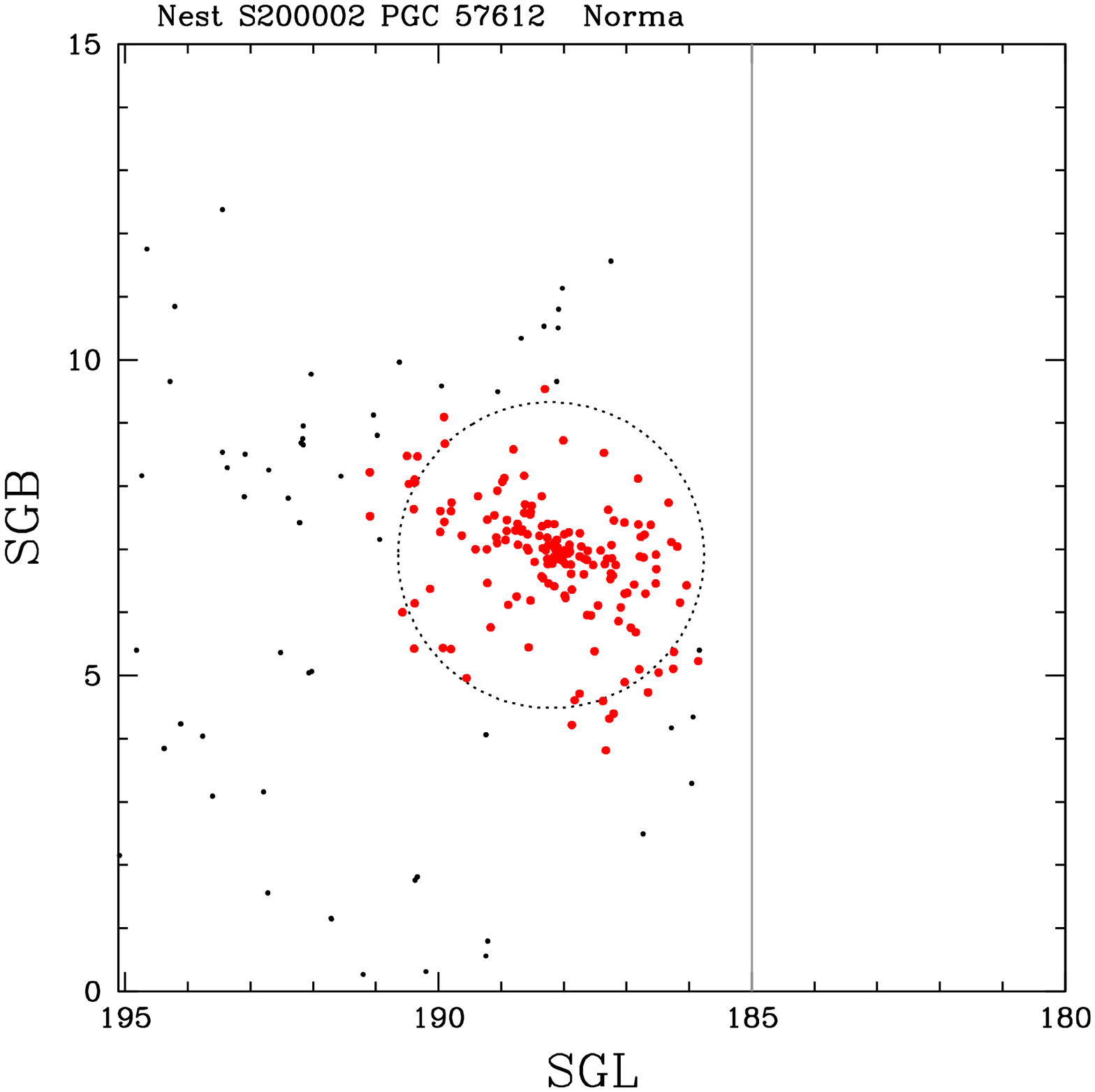}
\includegraphics[scale=0.4]{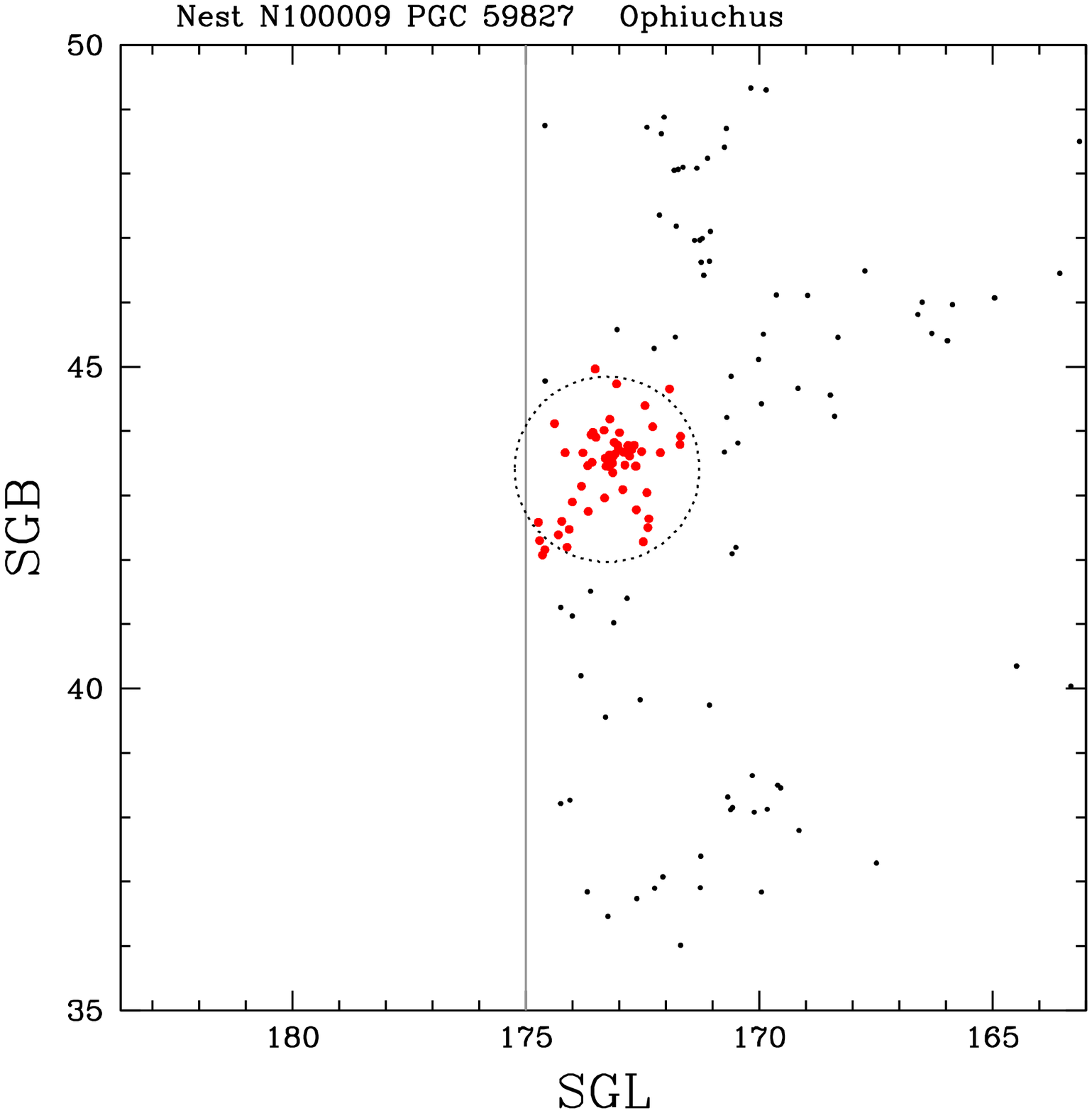}
\caption{Norma Cluster shown in the top panel and Ophiuchus Cluster shown in the bottom panel are two major clusters at low Galactic latitude.  The plane of the Milky Way lies at SGL=180 and the grey lines conform to the boundary of the 2MRS redshift survey. The PGC number identifies the dominant galaxy in the cluster.}
\label{norm-oph}
\end{center}
\end{figure}

Another detail is that the north and south Galactic hemispheres were evaluated separately since no real group except the most local crosses the zone of obscuration.  In the final group catalog, nests are given 6-figure identification numbers with the initial digit for those in the north the number 1 and the initial digit for those in the south the number 2.  The 2MRS limit of $\vert b \vert = 5^{\circ}$ fortuitously allows inclusion of two of the largest nearby clusters, Norma and Ophiuchus, as shown in Figure~\ref{norm-oph}.  The excluded low Galactic latitude zone occupies 9\% of the sky.  As an aside, a systematic X-ray search for clusters has revealed 3 clusters within 10,000~\kms\ in the 2MRS exclusion zone: CIZAJ0450.0+4501 ($V_h=6655$ \kms) and CIZAJ0603.8+2939 ($V_h=8994$ \kms) in proximity to the Perseus Cluster and CIZAJ1324.7-5736 ($V_h=5696$~\kms) near the Norma Cluster \citep{2002ApJ...580..774E}. 

\begin{figure}[htbp]
\begin{center}
\includegraphics[scale=0.4]{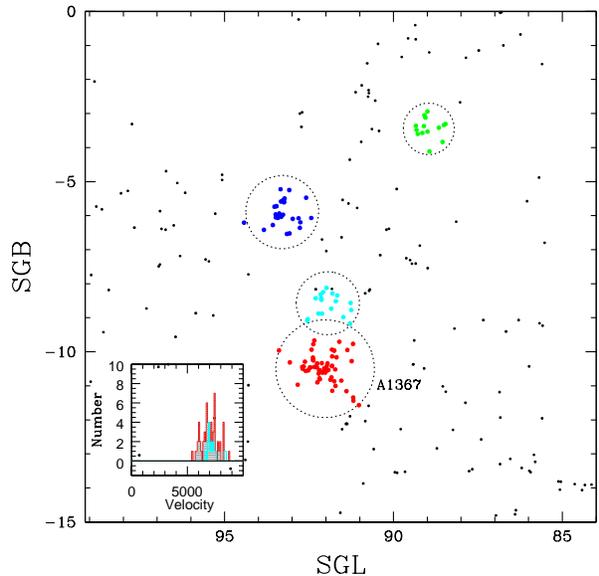}
\caption{The region around the Abell 1367 Cluster.  The main cluster in red and an apparently distinct entity in cyan have strongly overlapping velocities.}
\label{a1367}
\end{center}
\end{figure}

\begin{figure}[]
\begin{center}
\includegraphics[scale=0.4]{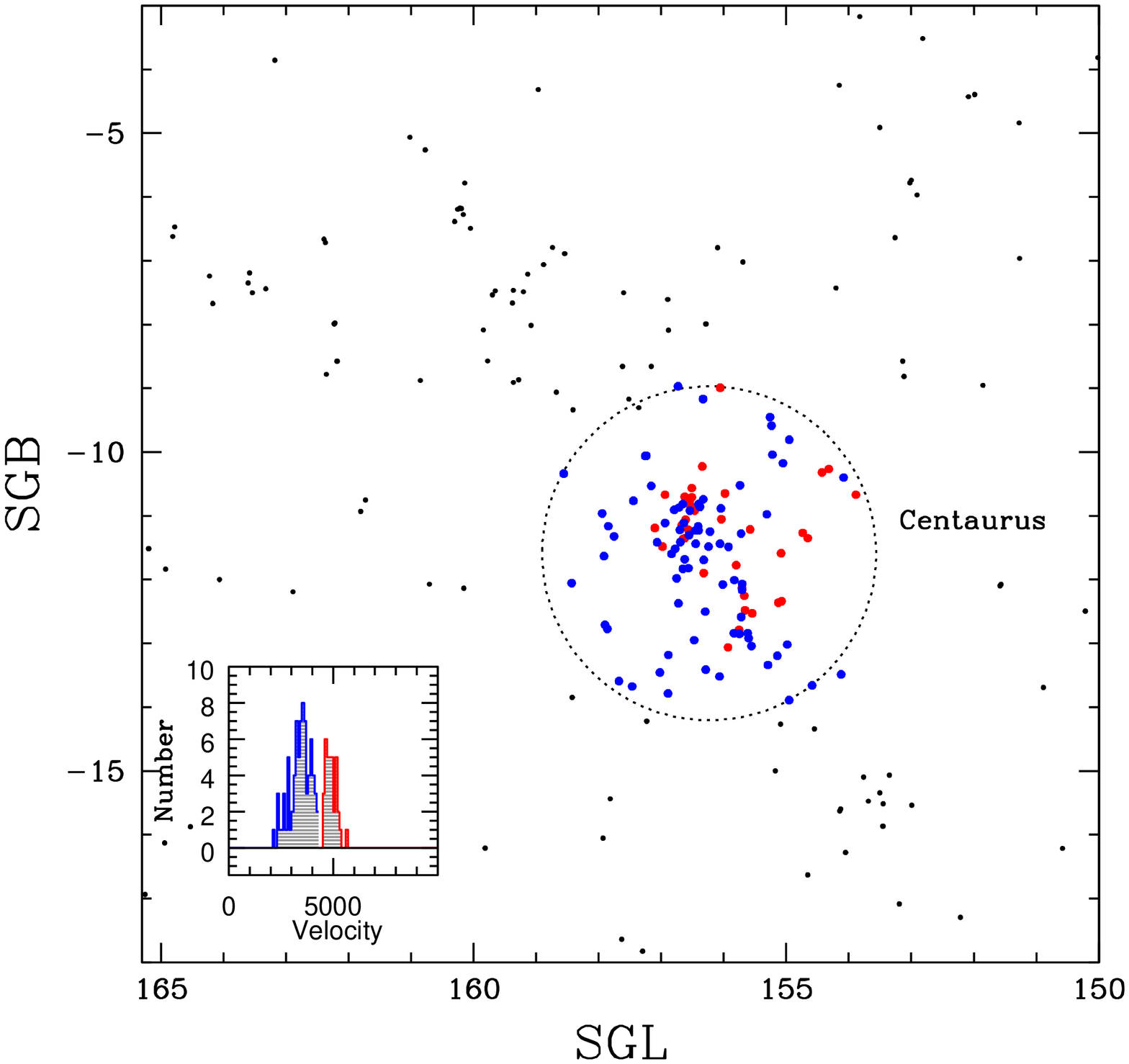}
\caption{The region around the Centaurus Cluster.  Two relatively distinct kinematic components strongly overlap in projection.  The blue and red components separated at 4300~\kms\ have been called Cen30 and Cen45 in the literature.}
\label{cen}
\end{center}
\end{figure}

A third, more significant detail follows from close inspection of the nests in plots of projected positions and histograms of velocities.  There sometimes were apparent gaps that suggest the linkages were too severe.  These gaps were usually more convincing in the spatial information rather than in velocities. In only six instances, nests were split.  In these cases the smaller part is given an identification with the number 2 as the second digit.   The digits 3 to 6 are the same as the identification number of the larger part.   The Abell 1367 Cluster seen in Figure~\ref{a1367} is an example.  The main cluster in red is nest 100005 while the secondary cluster in cyan is nest 120005 in the group catalog.  It can be expected in such a case that the two entities are on paths to soon merge.  Nests are split where the evidence is compelling.  With several other nests there are hints of segregation but there is sufficient ambiguity that they were left alone.  For example, in the case of the Centaurus Cluster shown in Figure~\ref{cen} there appear to be two kinematic components but they are not distinctly separated and the spatial overlap is severe.  The two components in this well known case \citep{1986MNRAS.221..453L, 1997A&A...327..952S} appear to have similar distances.  Here and commonly with the halos under consideration it can be expected that merging and cluster growth is an ongoing process.

A fourth detail results from the observation that the algorithm that is employed fails to pick up extreme velocity outliers.  With the large nests it was not uncommon to find velocity outliers reaching $\sim 3.5\sigma$ projected very near to the center of the nest.  Their central locations and the paucity of galaxies in the vicinity with similar velocities makes the likelihood of their membership very high.   These galaxies are added to the associated nests and the nest parameters are recalculated.

Table~\ref{summary} gives the gross statistics of the group catalog, both overall and in the restricted interval 3,000 to 10,000 \kms.  In the velocity interval 3,000~\kms\ to 10,000~\kms\ there are 24,044 galaxies to consider in the 2MRS 11.75 sample.  These are assembled into 13,605 groups, with 3,461 groups of at least 2 members and 10,144 singles.  The Perseus Cluster contains the most galaxies from the 2MRS 11.75 sample in this restricted velocity range, with 180.  Figure~\ref{cf} teaches us that number counts are strongly dependent on distance.  Masses from adjusted luminosities and velocity dispersions are more stable with distance.  Table~\ref{prominent} lists the top 8 clusters in the $3-10$ thousand velocity interval: on the left, by masses derived from luminosities, and on the right, by velocity dispersions.  For comparison, the Virgo Cluster, the most important cluster within 3,000~\kms\ is included.  The Virgo Cluster is a comfortable peer of these prominent groups.
 
\begin{deluxetable}{lcccccc}
\tablecaption{Summary}
\tablewidth{0pt}
\tablehead{
\colhead{Sample}&\colhead{\# All Galaxies}&\colhead{\# Groups to 20k}&\colhead{$\ge 2$}&\colhead{\# 3k$-$10k}&\colhead{\# Groups}&\colhead{$\ge 2$}
}
\startdata
North & 21,995 & 13,090 & 3,178 & 12,153 & 6,982 & 1,781 \\
South & 21,043 & 12,376 & 3,010 & 11,891 & 6,624 & 1,680 \\
Total  & 43,038 & 25,475 & 6,188 & 24,044 & 13,606 & 3,461 \\
\enddata
\label{summary}
\end{deluxetable}

\begin{deluxetable}{llccclc}
\tablecaption{Most Prominent Clusters}
\tablewidth{0pt}
\tablehead{
\colhead{Group}&\colhead{Name}&\colhead{Mass $ h^{-1}10^{15} \Msun$}&\colhead{CF}&\colhead{Group}&\colhead{Name}&\colhead{$\sigma_p$ \kms}
}
\startdata
100004 & A2199        & $1.8$ & 2.3 & 100009 & Ophiuchus & 976 \\
100001 & Coma         & $1.7$ & 1.6 & 200001 & Perseus      & 962 \\
100009 & Ophiuchus & $1.6$ & 2.3 & 200002 & Norma        & 957 \\
200002 & Norma        &  $1.3$ & 1.3 & 100001 & Coma         & 886  \\
200001 & Perseus      &  $1.2$ & 1.2 & 100003 & Centaurus & 822 \\
100002 & Virgo           &  $1.1$ & 1.0 & 200017 & A539           & 754 \\
200016 & A2634         &  $0.8$ & 2.3 & 100004 & A2199        & 740 \\
200017 & A539            & $0.7$ & 1.9 & 100002 & Virgo           & 717 \\
100005 & A1367         &  $0.7$ & 1.5 & 100005 & A1367         & 707 \\
\enddata
\label{prominent}
\end{deluxetable}

\subsection{Local Densities}
\label{sec:locden}

Knowledge of the selection function and of the group characteristics allows for the determination of the smoothed luminous density at the location of every galaxy in the sample.  Values for this parameter have been calculated on a $1 h^{-1}$ Mpc grid and each galaxy is given the value at the nearest grid location.  The luminosity densities at each grid point are given by the sum of contributions from members of the sample after Gaussian smoothing.  Luminosities are adjusted following the recipe illustrated in Fig.~\ref{cf}.  The Gaussian smoothing scale is $1h^{-1} \times CF_{lum}^{1/3}$~Mpc so the compensation for missing light is spread across more grid points at larger distances while the contribution from a single galaxy remains constant at the central position of the galaxy.  In order to eliminate the finger-of-god elongation of structures in redshift space, yet realistically replicate the depth of clusters, galaxies in the identified groups with velocity $V_{gal}$ are given redshift space positions in the line-of-sight based on the mean group velocity $V_{gp}$ plus $0.1 (V_{gal}-V_{gp})$.   Finally to note: since the luminosity adjustment becomes very large at large redshifts, the luminosity density computation is restricted to a box $\pm 200 h^{-1}$ Mpc on a side.  Even within this box, density values toward the edges are very uncertain.

\begin{figure}[]
\begin{center}
\includegraphics[scale=0.32]{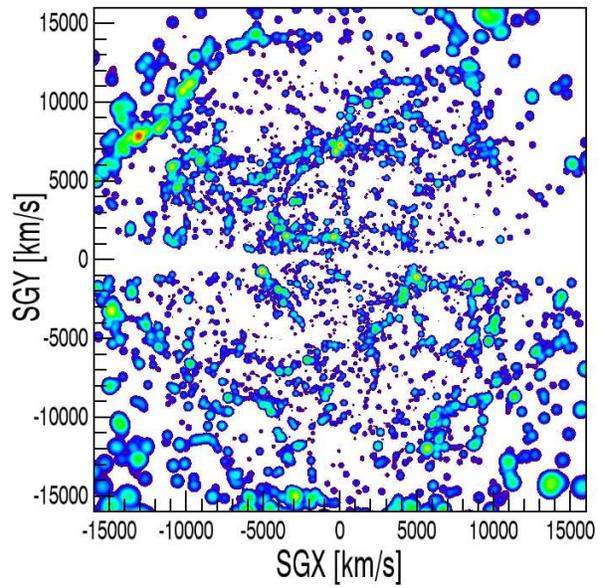}
\caption{Slice through smoothed 2MRS luminosity density cube: $SGZ$ interval -2000 to +2000~\kms.  Contours are at logarithmic intervals with a low density cutoff at $10^9~h^2~\Msun~{\rm Mpc}^{-3}$. Galactic obscuration band at $SGY \sim 0$. The most prominent feature is the Shapley Concentration at $SGX=-13,000$~\kms, $SGY=8,000$~\kms.}
\label{2mrsdens_Z}
\end{center}
\end{figure}

\begin{figure}[]
\begin{center}
\includegraphics[scale=0.32]{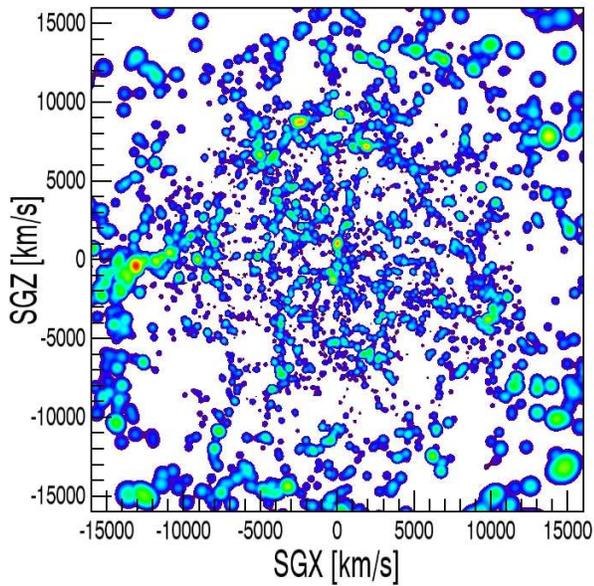}
\caption{Slice through smoothed 2MRS luminosity density cube: $SGY$ interval +5000 to +9000~\kms.  Contours as in Fig.~\ref{2mrsdens_Z}.  This interval contains the Great Wall with the Coma Cluster near the center, the Hercules complex above center and slightly to the left, and the Shapley Concentration at $SGX=-13,000$~\kms, $SGZ=0$.}
\label{2mrsdens_Y}
\end{center}
\end{figure}

\begin{figure}[]
\begin{center}
\includegraphics[scale=0.32]{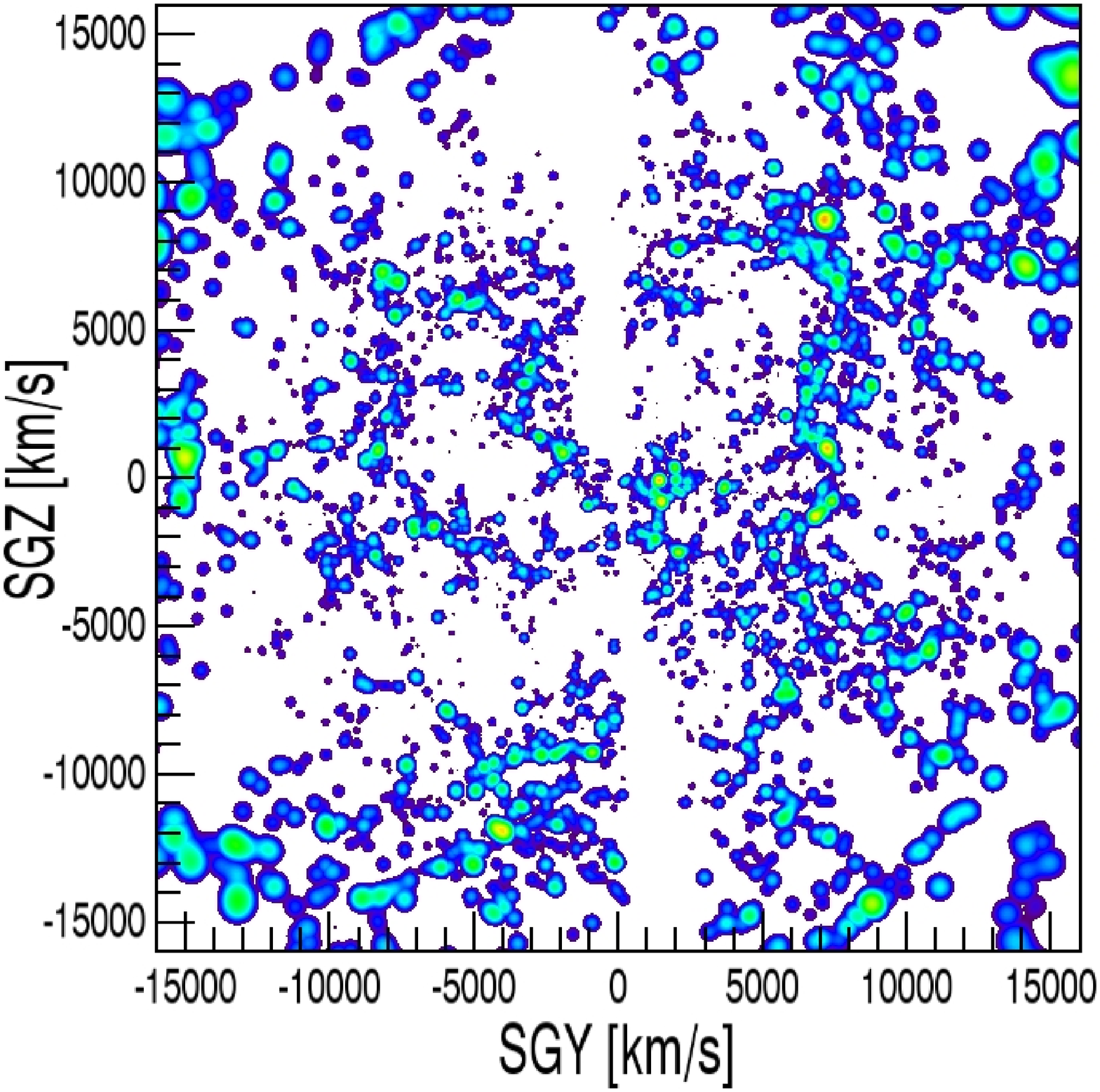}
\caption{Slice through smoothed 2MRS luminosity density cube: $SGX$ interval -4000 to 0~\kms. Contours as in Fig.~\ref{2mrsdens_Z}.  Galactic obscuration is tilted by $6^{\circ}$ from the $SGY=0$ axis.  The Great Wall rises vertically at $SGY=6,000$~\kms\ reaching the Hercules complex at $SGZ+8,000$~\kms.  Directly above our position at the origin is the Local Void, blending toward the upper right into the Hercules Void.}
\label{2mrsdens_X}
\end{center}
\end{figure}

Maps of the density distribution are shown in three orthogonal slices in 
Figures~\ref{2mrsdens_Z}, ~\ref{2mrsdens_Y}, and \ref{2mrsdens_X}.
The presentations are in supergalactic coordinates and each slice has a thickness of 4000~\kms.  The slices are chosen to include most of the main structural features in the volume being studied.  See the captions for details.

\subsection{Tabulated Information}

Three on-line tables are provided that present the group catalog.  Table~\ref{nests} gives a summary of each group, one line per group.  Table~\ref{members} identifies the individual members of each group.  Table~\ref{eddcat} combines these pieces of information: there is one line per galaxy containing information about the galaxy and about the ensemble properties of its group.  In each case, the first 10 rows of large tables are shown, with the full tables made available on-line. Table \ref{eddcat} is also available (separated into Galactic north and south halves) at the Extragalactic Distance Database.\footnote{http://edd.ifa.hawaii.edu; see catalogs `2MRS North \& South Groups'}  More detailed information about the tables follows. 

\noindent
{\it Table \ref{nests}: Summary of Group Properties.}  Column (1) Group (nest) identification number; 1xxxxxx if in north galactic hemisphere and 2xxxxxx if in south. If the identification is 12xxxxx or 22xxxxx the group is a split from an adjacent larger group.  (2) Number of members from the 2MASS 11.75 catalog. (3) Principal Galaxies Catalog (PGC) name for brightest member in group. (4$-$5) Luminosity weighted supergalactic longitude and latitude of group, in degrees.  (6) Logarithm of $K_s$ band luminosity summed over all 2MASS 11.75 members and adjusted by the correction factor for missing light, units of solar luminosity.  (7) Unweighted average group velocity, in the cosmic microwave background frame adjusted by a cosmological model as described in \cite{2013AJ....146...86T}, in \kms. (8) Distance modulus from group velocity assuming H$_0=100$~\kmsMpc, in magnitudes. (9) Line of sight bi-weight velocity dispersion of group members; requires at least 5 group members, in \kms. (10) Projected radius of second turnaround calculated from group luminosity and assumed conversion to mass, at group distance given by velocity and H$_0=100$~\kmsMpc, in Mpc. (11) Line of sight velocity dispersion anticipated from group luminosity and assumed conversion to mass, in \kms.  (12) Group mass calculated from luminosity, adjusted by correction factor, and assumed light to mass conversion factor.  (13) Conversion factor for missing light as a function of systemic velocity.

\noindent
{\it Table \ref{members}: Group Members.}  Column (1) Group (nest) identification number; 1xxxxxx if in north galactic hemisphere and 2xxxxxx if in south. If the identification is 12xxxxx or 22xxxxx the group is a split from an adjacent larger group. (2) Principal Galaxies Catalog (PGC) name of group member.  (3$-$4) Supergalactic longitude and latitude of galaxy, in degrees. (5) Type in the numeric code of \citet{1991trcb.book.....D}. (6) 2MASS $K_s$ total magnitude, extinction corrected, in magnitudes. (7) 2MASS $J-H$ total extinction corrected color, in magnitudes, (8) 2MASS $J-K$ total extinction corrected color, in magnitudes.  (9) Logarithm of intrinsic $K_s$ luminosity, extinction corrected, at group distance given by velocity and H$_0=100$~\kmsMpc, in units of solar luminosity. (10) Velocity of galaxy in the cosmic microwave background frame adjusted by a cosmological model as described in \cite{2013AJ....146...86T}, in \kms.

\noindent
{\it Table \ref{eddcat}: All Galaxies with Group Tags.}\footnote{Available with updates at http://edd.ifa.hawaii.edu} Column (1) Principal Galaxies Catalog (PGC) name of galaxy. (2$-$3) Galactic longitude and latitude, in degrees. (4$-$5) Supergalactic longitude and latitude, in degrees. (6) Type in the numeric code of \citet{1991trcb.book.....D}. (7) Heliocentric velocity, in \kms. (8) Velocity in the Local Sheet reference frame defined by \cite{2008ApJ...676..184T}, in \kms. (9) Velocity of galaxy in the cosmic microwave background frame adjusted by a cosmological model as described in \cite{2013AJ....146...86T}, in \kms. (10) 2MASS $J-H$ total extinction corrected color, in magnitudes, (11) 2MASS $J-K_s$ total extinction corrected color, in magnitudes. (12) 2MASS $K_s$ total magnitude, extinction corrected, in magnitudes. (13) Logarithm of intrinsic $K_s$ luminosity, extinction corrected, at group distance given by velocity and H$_0=100$~\kmsMpc, in units of solar luminosity. (14) Logarithm of luminosity density at nearest location on a $1h^{-1}$ Mpc grid, where contributions come from all adjacent 2MASS 11.75 galaxies after smoothing with a Gaussian of $1h^{-1}$ Mpc times an adjustment dependent on the luminosity correction factor as described in the text, in units of solar luminosity per Mpc$^3$.  (15)  Group (nest) identification number; 1xxxxxx if in north galactic hemisphere and 2xxxxxx if in south. If the identification is 12xxxxx or 22xxxxx the group is a split from an adjacent larger group. (16) Number of members from the 2MASS 11.75 catalog. (17) Principal Galaxies Catalog (PGC) name for brightest member in group. (18$-$19) Distance and distance modulus from group velocity assuming H$_0=100$~\kmsMpc, in Mpc and magnitudes respectively. (20$-$21) Luminosity weighted supergalactic longitude and latitude of group, in degrees. (22) Logarithm of $K_s$ band luminosity summed over all 2MASS 11.75 members and adjusted by the correction factor for missing light, units of solar luminosity. (23) Conversion factor for missing light as a function of systemic velocity. (24) Line of sight velocity dispersion anticipated from group luminosity and assumed conversion to mass, in \kms. (25) Projected radius of second turnaround calculated from group luminosity and assumed conversion to mass, at group distance given by velocity and H$_0=100$~\kmsMpc, in Mpc. (26) Unweighted average group velocity, in the cosmic microwave background frame adjusted by a cosmological model as described in \cite{2013AJ....146...86T}, in \kms. (27) Bi-weight group \citep{1990AJ....100...32B} velocity, in the cosmic microwave background frame adjusted by the cosmological model, in \kms. (28) Uncertainty in bi-weight group velocity, in \kms. (29) Line of sight bi-weight velocity dispersion of group members; requires at least 5 group members, in \kms. (30) Line of sight velocity dispersion of group members, in \kms.  Null if group of one. (31) Bi-weight projected gravitational radius $R_{ij}$, in Mpc. (32) Uncertainty in bi-weight projected gravitational radius, in Mpc. (33) Group mass from virial theorem with bi-weight dispersion and radius parameters, in units of $10^{12}~\Msun$. (34) Group mass based on adjusted intrinsic luminosity and mass to light prescription, in units of $10^{12}~\Msun$. (35$-$36) \citet{2007ApJ...655..790C} high and low density group identifications. (37)  \citet{2011MNRAS.416.2840L} group identification. (38$-$40) Supergalactic X,Y,Z coordinates from group velocities assuming H$_0=100$~\kmsMpc, with small adjustments described in section~\ref{sec:locden} to create roughly spherical clusters, in Mpc.

\begin{deluxetable}{crrrrcrcrcrcr}
\tablecaption{Nest Properties (1st 10 of 25474)}
\tablewidth{0pt}
\tablehead{
\colhead{Nest}&\colhead{Mem}&\colhead{PGC1}&\colhead{$SGL$}&\colhead{$SGB$}&\colhead{log$L_K^g$}&\colhead{$V_{mod}^g$}&\colhead{DM}&\colhead{$\sigma_V$}&\colhead{$R_{2t}$}&\colhead{$\sigma_p$}&\colhead{Mass}&\colhead{$CF$}
}
\startdata
100001 & 136 &  44715 &  89.6226 &   8.1461 & 13.15 &  7331 & 34.33 &  886 & 2.393 &  881 & 0.171E+16 &   1.65\\
100002 & 197 &  41220 & 103.0008 &  -2.3248 & 12.69 &  1491 & 30.87 &  670 & 1.617 &  596 & 0.529E+15 &   1.00\\
100003 & 113 &  43296 & 156.2336 & -11.5868 & 12.75 &  3873 & 32.94 &  822 & 1.708 &  629 & 0.623E+15 &   1.12\\
100004 &  81 &  58265 &  71.5103 &  49.7851 & 13.16 &  9424 & 34.87 &  740 & 2.418 &  891 & 0.177E+16 &   2.26\\
100005 &  61 &  36487 &  92.0255 & -10.4950 & 12.78 &  6987 & 34.22 &  707 & 1.753 &  646 & 0.673E+15 &   1.58\\
100006 &  85 &  31478 & 139.4478 & -37.6063 & 12.48 &  4099 & 33.06 &  648 & 1.347 &  496 & 0.305E+15 &   1.14\\
100007 &  86 &  56962 & 108.5182 &  49.0878 & 13.52 & 11603 & 35.32 & 1261 & 3.168 & 1167 & 0.398E+16 &   3.30\\
100008 &  65 &  39600 &  67.2953 &   3.2390 & 11.94 &  1054 & 30.11 &  209 & 0.837 &  308 & 0.733E+14 &   1.00\\
100009 &  66 &  59827 & 173.2412 &  43.4150 & 13.13 &  9112 & 34.80 &  976 & 2.352 &  866 & 0.163E+16 &   2.15\\
100010 &  55 &  47202 & 149.1678 &  -1.3439 & 13.93 & 15265 & 35.92 & 1002 & 4.363 & 1607 & 0.104E+17 &   7.95\\
\enddata
\label{nests}
\end{deluxetable}

\begin{deluxetable}{crrrrrccrr}
\tablecaption{Nest Members (1st 10 of 43038)}
\tablewidth{0pt}
\tablehead{
\colhead{Nest}&\colhead{PGC}&\colhead{$SGL$}&\colhead{$SGB$}&\colhead{Ty}&\colhead{$K_{tot}$}&\colhead{$J-H$}&\colhead{$J-K$}&\colhead{log$L_K$}&\colhead{$V_{mod}$}
}
\startdata
100001 &   44715 &  89.6369 &   8.3951 & -4.3 &  8.40 &  0.69 &  0.99 & 11.69 &  6834\\
100001 &   44628 &  89.6262 &   8.2751 & -3.6 &  8.86 &  0.82 &  0.98 & 11.50 &  7582\\
100001 &   43895 &  90.2180 &   6.9260 & -2.3 &  9.24 &  0.47 &  0.75 & 11.35 &  8837\\
100001 &   44298 &  89.9643 &   7.6957 & -3.9 &  9.19 &  0.75 &  0.88 & 11.37 &  7763\\
100001 &   44323 &  89.0106 &   7.9544 & -4.1 &  9.22 &  0.81 &  1.03 & 11.36 &  7164\\
100001 &   44840 &  89.8613 &   8.5241 &  4.0 &  9.84 &  0.67 &  0.98 & 11.11 &  8420\\
100001 &   44938 &  90.0745 &   8.6928 & -3.1 &  9.89 &  0.67 &  0.94 & 11.09 &  8300\\
100001 &   43981 &  89.9140 &   7.1398 & -2.6 &  9.89 &  0.70 &  1.02 & 11.09 &  8288\\
100001 &   44737 &  89.4237 &   8.4824 & -2.1 & 10.14 &  0.68 &  0.94 & 10.99 &  8998\\
100001 &   44114 &  89.6574 &   7.4958 & -3.0 &  9.71 &  0.68 &  0.98 & 11.16 &  7323\\
\enddata
\label{members}
\end{deluxetable}

\begin{deluxetable}{rrrrrrrrrccrrrcrrrrrrrrrcrrrrrccrrrrrrrr}
\tablecaption{Combined Catalog (1st 10 of 43038)}
\tablewidth{0pt}
\tablehead{
\colhead{PGC}&\colhead{Glon}&\colhead{Glat}&\colhead{$SGL$}&\colhead{$SGB$}&\colhead{Ty}&\colhead{$V_{hel}$}&\colhead{$V_{LS}$}&\colhead{$V_{mod}$}&\colhead{$J-H$}&\colhead{$J-K$}&\colhead{$K_{tot}$}&\colhead{log$L_K$}&\colhead{log$\rho_K$}&\colhead{Nest}&\colhead{Ng}&\colhead{PGC1}&\colhead{$D_V$}&\colhead{DM}&\colhead{$SGL_g$}&\colhead{$SGB_g$}&\colhead{log$L_K^g$}&\colhead{$CF$}&\colhead{$\sigma_p$}&\colhead{$R_{2t}$}&\colhead{$V_mod^g$}&\colhead{$V_{bw}^g$}&\colhead{$eV_{bw}$}&\colhead{$\sigma_{bw}$}&\colhead{$V_{rms}$}&\colhead{$R_{ij}^{bw}$}&\colhead{$eR_{ij}$}&\colhead{$M_{12}^{vir}$}&\colhead{$M_{12}^{lum}$}&\colhead{HDC}&\colhead{LDC}&\colhead{2M++}&\colhead{$SGX$}&\colhead{$SGY$}&\colhead{$SGZ$}
}
\startdata
      2 & 113.9554 & -14.6992 & 341.6442 &  20.7389 &  3.1 &  5017 &  5309 &  4799 &  0.72 &  1.04 &  9.50 & 10.91 & 10.41 & 200275 &   7 &   73150 &  49.8 & 33.48 & 341.4922 &  20.7395 & 11.57 &   1.24 &  222 & 0.604 &  4976 &  4984 &  60 &  159 &  155 & 0.673 & 0.070 &    18.600 &    27.500 &   11 &    2 & 4634 &   44.20 &  -14.67 &   17.63\\
      5 & 110.6206 & -28.9043 & 326.1772 &  19.7805 & -0.6 & 10445 & 10717 & 10310 &  0.70 &  1.01 & 10.49 & 11.12 & 10.34 & 200619 &   4 &       5 & 100.7 & 35.02 & 326.1907 &  19.7757 & 12.01 &   2.56 &  328 & 0.890 & 10075 &       &     &      &  276 &       &       &           &    88.100 &    1 &    1 &    0 &   78.72 &  -52.75 &   34.08\\
     12 &  90.1920 & -65.9300 & 286.4249 &  11.3510 &  1.3 &  6546 &  6683 &  6279 &  0.72 &  0.99 & 11.11 & 10.46 &  9.36 & 210177 &   1 &      12 &  62.8 & 33.99 & 286.4249 &  11.3510 & 10.64 &   1.49 &   98 & 0.266 &  6279 &       &     &      &    0 &       &       &           &     2.340 &    0 &    0 &    0 &   17.40 &  -59.06 &   12.36\\
     14 & 101.7853 & -52.4728 & 300.9143 &  15.3673 & -4.9 & 11602 & 11801 & 11571 &  0.76 &  1.03 & 10.60 & 11.22 & 10.57 & 200601 &   4 &      14 & 118.2 & 35.36 & 300.8070 &  15.3909 & 12.22 &   3.61 &  395 & 1.071 & 11820 &       &     &      &  311 &       &       &           &   154.000 &    0 &    0 &    0 &   58.55 &  -97.79 &   31.32\\
     16 &  91.6006 & -64.8655 & 287.6120 &  11.7030 & -0.1 &  5664 &  5806 &  5434 &  0.63 &  0.82 & 11.52 & 10.18 &  9.78 & 211419 &   1 &      16 &  54.3 & 33.68 & 287.6120 &  11.7030 & 10.31 &   1.34 &   73 & 0.198 &  5434 &       &     &      &    0 &       &       &           &     0.977 &    0 &    0 &    0 &   16.08 &  -50.68 &   11.01\\
     18 & 113.9223 & -15.0086 & 341.3140 &  20.7012 &  2.0 &  5366 &  5658 &  5156 &  0.66 &  0.84 & 10.69 & 10.43 & 10.41 & 200275 &   7 &   73150 &  49.8 & 33.48 & 341.4922 &  20.7395 & 11.57 &   1.24 &  222 & 0.604 &  4976 &  4984 &  60 &  159 &  155 & 0.673 & 0.070 &    18.600 &    27.500 &   11 &    2 & 4634 &   44.13 &  -14.93 &   17.60\\
     23 &  94.2541 & -62.6047 & 290.1153 &  12.4061 & -2.8 & 11365 & 11518 & 11326 &  0.78 &  0.68 & 10.94 & 11.04 & 10.12 & 202425 &   2 &      23 & 112.8 & 35.26 & 290.0671 &  12.6568 & 11.85 &   3.27 &  285 & 0.773 & 11282 &       &     &      &   62 &       &       &           &    57.700 &    0 &    0 &    0 &   37.88 & -103.45 &   24.23\\
     25 & 107.8762 & -38.3822 & 315.9687 &  18.3751 & -3.0 & 11691 & 11939 & 11700 &  0.69 &  0.99 & 10.97 & 11.06 &  9.84 & 205376 &   1 &      25 & 117.0 & 35.34 & 315.9687 &  18.3751 & 11.61 &   3.48 &  230 & 0.625 & 11700 &       &     &      &    0 &       &       &           &    30.500 &    0 &    0 &    0 &   79.82 &  -77.18 &   36.88\\
     30 & 305.4397 & -36.0852 & 215.6453 & -13.3490 &  4.1 &  7907 &  7691 &  8058 &  0.60 &  0.96 &  9.98 & 11.13 &  9.93 & 206399 &   1 &      30 &  80.6 & 34.53 & 215.6453 & -13.3490 & 11.37 &   1.89 &  186 & 0.506 &  8058 &       &     &      &    0 &       &       &           &    16.200 &    0 &    0 &    0 &  -63.73 &  -45.70 &  -18.61\\
     31 & 326.3370 & -67.7221 & 247.9873 &  -2.4720 & 99.0 &  6032 &  5965 &  5867 &  0.63 &  0.99 & 10.46 & 10.66 &  9.85 & 202464 &   2 &      31 &  58.3 & 33.83 & 247.8761 &  -2.5482 & 11.01 &   1.41 &  136 & 0.368 &  5827 &       &     &      &   57 &       &       &           &     6.240 &    0 &    0 &    0 &  -21.84 &  -54.00 &   -2.51\\
\enddata
\label{eddcat}
\end{deluxetable}

\section{Group Properties}
\label{sec:gpprop}

It can be asked if observed velocity dispersion, spatial scale, and inferred virial masses track the assumptions based on luminosity.  Correlations are expected but there could be offsets.  There will be evident uncertainties with small groups.  In best cases with order 100 candidate members, projected velocities and positions reasonably represent the three-dimensional distribution but with only a few members, in the extreme only two, knowledge of only one of three velocity components and two of three spatial components results in large uncertainties in group properties.  Uncertainties are compounded by interlopers.  Then, more fundamental than the observational considerations, it is appreciated that the structures may often, even usually, stray from dynamical equilibrium. 

While ultimately uncertainty will rule, there are better and worse ways to evaluate group parameters.  Approaches have been discussed by \citet{1990AJ....100...32B}.  They evaluate methods for what they call resistance, robustness, and efficiency.  A resistant method is minimally affected by outliers.  A robust method works with diverse population characteristics.  An efficient method does the best that can be done with poor statistics.  Following tests with samples from 5 to 200 in size, Beers et al. identify a preference for bi-weight location and scale estimators.  Their recipes are followed here to determine group central velocities and line-of-sight dispersions and measures of the projected separations $R_{ij}$ of group members.

\begin{figure}[htbp]
\begin{center}
\includegraphics[scale=0.4]{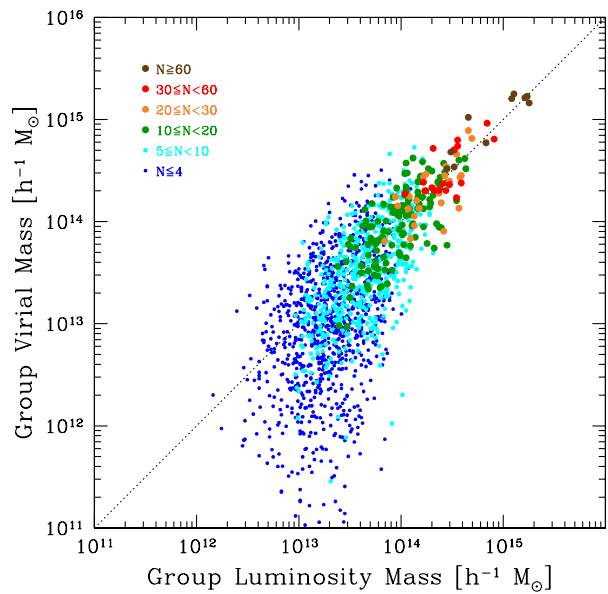}
\caption{Comparison of group mass estimates based on luminosity vs. estimates based on velocity dispersions and spatial separations.  The color code identifies the number of members.}
\label{MlMv}
\end{center}
\end{figure}

Figure~\ref{MlMv} provides a comparison between group mass estimates derived via the virial theorem from positions and velocities versus group mass estimates following from the integrated $K_s$ band light.  The dotted line gives the equality relationship.  It is seen that there is excellent agreement with groups containing at least 30 2MRS galaxies, identified by the brown and red points,  and still good agreement with memberships as small as 10, points in orange and green.  The correlation falls apart with fewer group members, already with 5 to 9 members represented in the plot in cyan, but especially with groups of 2 to 4 represented in blue.  

Of course, the virial mass estimates are suspect with small numbers.  The systematic trend toward small masses can be understood.  The group search algorithm  that sets spatial and velocity limits favors inclusion of galaxies with close projected separations and radial velocities and disfavors objects with large projected separations and radial velocities that would make the cut with full 3D information.  The scaling laws assume that the accessible phase space information reasonably represents the group which is statistically valid if the membership is large.

\begin{figure}[htbp]
\begin{center}
\includegraphics[scale=0.4]{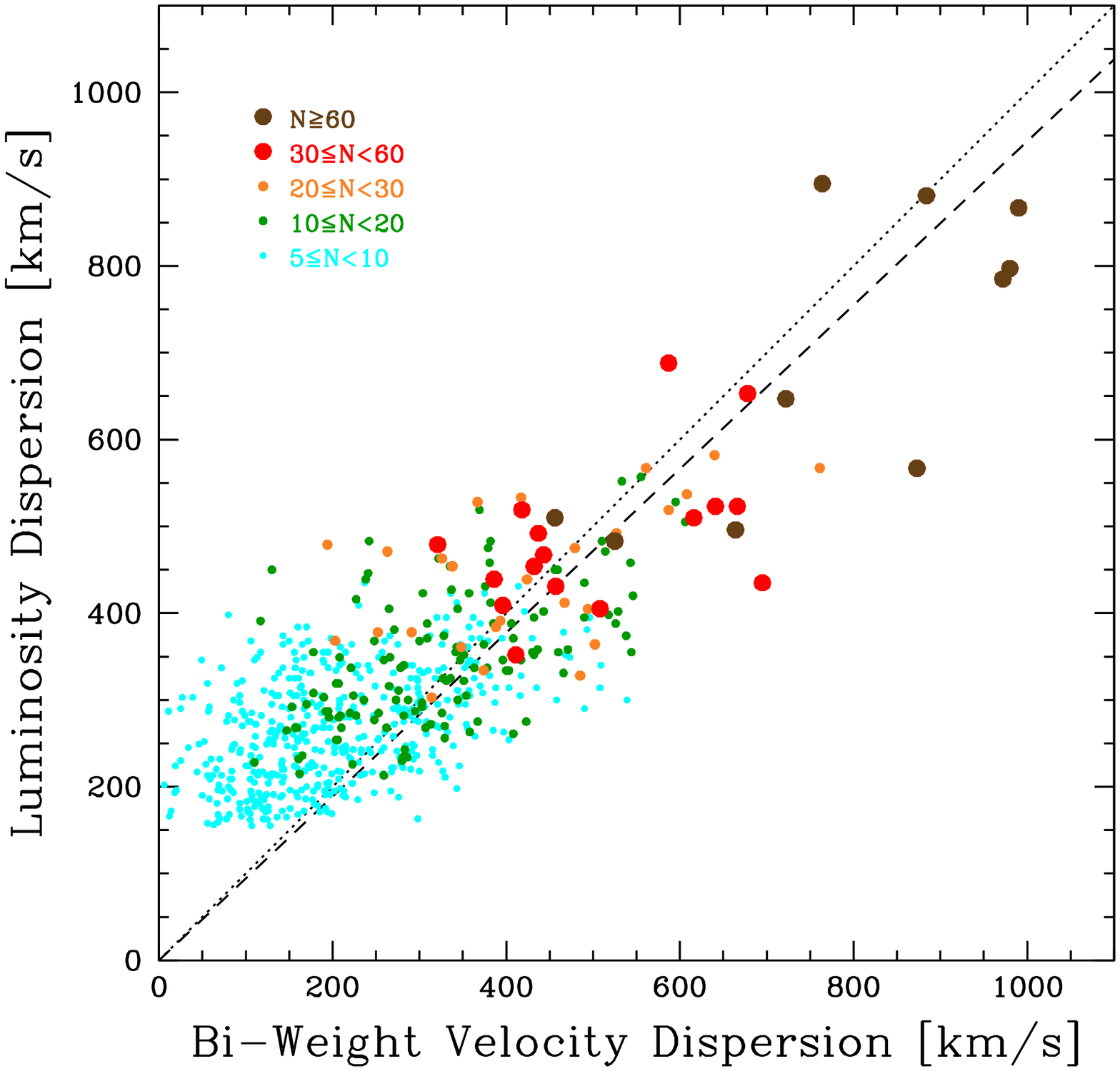}
\includegraphics[scale=0.4]{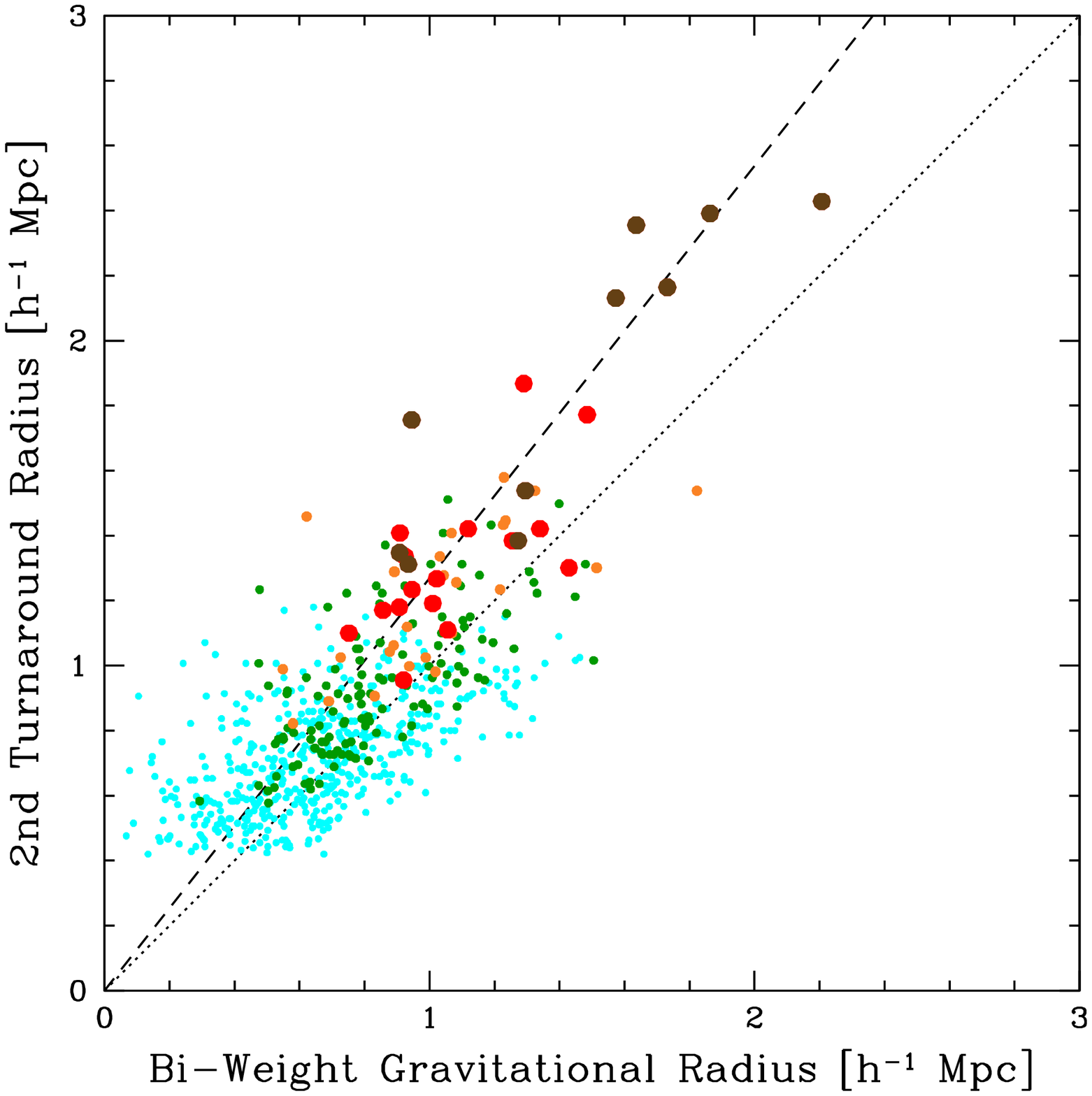}
\caption{{\it Top:} bi-weight radial velocity dispersion vs. velocity dispersion inferred from $K_s$ luminosity. {\it Bottom:} bi-weight projected gravitational radius vs. $R_{2t}$.
The color code identifies the number of members.  Dotted curves correspond to 1:1 dependencies.  Dashed curves are the best fits to groups with at least 30 members.}
\label{r_sig}
\end{center}
\end{figure}

The separate velocity dispersion and spatial scale components are given attention in Figure~\ref{r_sig}.  Observed dispersions are compared with dispersions anticipated from luminosities in the top panel and observed gravitational radii are compared with second turnaround radii in the bottom panel.  Color codings tied to group membership  are the same as in Fig.~\ref{MlMv} except groups with less than 5 members are omitted.  The absolute luminosity cutoff $M_{K_s}=-19+5{\rm log}h$ translates to lower cutoffs in the dispersions and dimensions derived from luminosities.  The dashed lines in the two panels represent best fits to the comparisons involving groups with at least 30 members.  In the case of velocity dispersions the difference between measures is marginal.  In the case of dimensions, there is no expectation that the two measures would be the same.  Statistically, $R_{2t}$ is $27\%\pm3\%$ larger than the gravitational radius. 

There are two conclusions from the discussion of this section.  The first is that the absolute scale of the masses derived from luminosities is consistent with a scale from kinematic input.  The second is that $K_s$ luminosities provide masses with smaller uncertainties than kinematic masses.  Admittedly there could be hidden systematics in the luminosity masses that remain to be revealed. 

\section{The Group Mass Function}

The availability of the group catalog presents the opportunity to construct a well formulated halo mass function.  The analysis is restricted to groups with mean velocities in the range 3,000$-$10,000~\kms\ in the CMB frame.  The definition of groups includes groups of one, so the concepts of groups and halos are synonymous to the degree that halos are filled with enough stars to be visible. Within the completion limits of the 2MRS 11.75 survey, the group catalog contains {\it all} halos in the volume.

\begin{figure}[!]
\begin{center}
\includegraphics[scale=0.4]{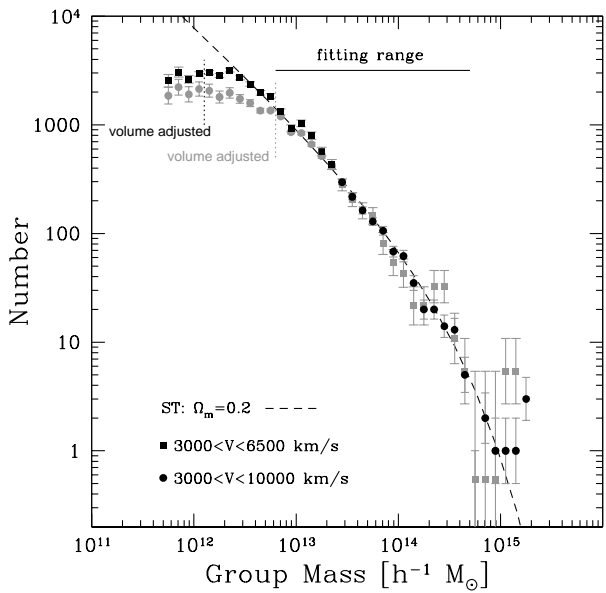}
\caption{Group mass function.  Corrections are made for loss of information with distance.  At masses above $3h^{-1}\times10^{13}~\Msun$ the mass function is built with contributions over the full range 3,000$-$10,000~\kms\ while at lower masses the construction is limited to contributions over the range 3,000$-$6,500~\kms.  The model fit is based on the Sheth-Tormen (ST) formalism with $\Omega_{matter}=0.2$ and a count normalization.}
\label{MN}
\end{center}
\end{figure}

The mass function derived with this sample is presented in Figure~\ref{MN}.  Masses are determined from $K_s$ luminosities, as justified in Section~\ref{sec:gpprop}.  It is seen that the mass function is built by merging two regimes.  The highest mass (luminosity) groups are seen across the full range of velocities (distances).  However high mass groups are rare in the nearer part of the survey because of the limited volume.  By contrast, the lowest mass groups are not seen in the outer shells but within the range that they can be seen they are present in substantial numbers. 

In Fig.~\ref{MN} there are overlays of the mass functions built in the two regimes: circles show the mass function with contributions from the full range 3,000$-$10,000~\kms\ and squares show the mass function restricted to 3,000$-$6,500~\kms.  The numbers for the shallower sample are geometrically augmented to compensate for the reduced volume being explored.  The two realizations of the mass function overlap satisfactorily in the mid range of $10^{13} - 10^{14}~\Msun$.  In the plot, circles are black and squares are grey above $3h^{-1} \times 10^{13}~\Msun$ while circles are grey and squares are black below this mass.  At the high mass end the contributions from the two regimes are consistent but the bin $\sqrt{N}$ error bars are larger for the shallower sample.  At the low mass end there are differences attributable to completion effects. 

The dashed curve in Fig.~\ref{MN} illustrates a fit\footnote{Kindly provided by Julien Carron and Melody Wolk.} using the Sheth$-$Tormen modification \citep{1999MNRAS.308..119S} of the Press$-$Schechter  formulation of the mass spectrum of collapsed halos from an initial Gaussian field \citep{1974ApJ...187..425P}.  Fits with recipes by \citet{2001MNRAS.321..372J} or \citet{2006ApJ...646..881W} are not significantly different from the Sheth$-$Tormen case.  The Sheth$-$Tormen and related descriptions of the mass function are motivated by Lambda Cold Dark Matter simulations.  

The model fit minimizes scatter in bins of log(Number) over the solar mass range 12.8 to 14.7 in log units, with all bins in that range given equal weight.  The fit excludes halos with log mass greater than 14.7 because of poor statistics and halos below log mass 12.8 because of probable systematics.  Over the fitted range, the scatter of 0.07 in log counts is achieved with a low value of $\Omega_{matter}\sim 0.2$, assuming  a topologically flat universe.  However, while the shape of the model fit is in excellent agreement with the data, the count normalization is poor.  There will be cosmic variance.  As a measure of statistical fluctuations, there are 7\% more 2MRS galaxies in the sample volume in the north Galactic hemisphere than in the same volume in the south.   However, the nominal Sheth$-$Tormen fit required an augmentation in counts by a factor 4.6 to achieve the fit shown in Fig.~\ref{MN}.  This unsatisfactory situation requires an investigation that goes beyond the scope of this paper for reasons discussed next.


The fit clearly fails at group masses below $6h^{-1} \times 10^{12}~\Msun$.  There are several possible reasons.  There is immediate concern related to the luminosity corrections with distance.  The adjustment reaches a factor two at 9,000~\kms\ and, in the extreme, the adjustment can be as large as 2.6 or a displacement of 4 bins in log mass.  Certainly an adjustment is required.  Adjustments cause a binning uncertainty at the level of $\pm1$ bin.  A related but greater concern is whether the correction for missing light should strictly be assigned to one halo.  The assumption to do so is probably valid at high masses but not valid at low masses.  High mass groups contain a lot of galaxies and statistically the galaxies missed will be associated.  Low mass groups contain few galaxies, possibly only one that makes the 2MRS 11.75 catalog.  It is likely that frequently all or part of the added luminosity lies in separate uncatalogued halos.  The consequence would be that the present procedure preserves total light but assigns the light to fewer halos than actually exist.  There would be depletion of the numbers in the lowest mass bins and possible augmentation of numbers in intermediate mass bins.  

One wonders about the effect of the loss of low surface brightness galaxies in the 2MASS samples.  The problem is not so much with the systematic loss of light.  If that light had been captured then the coefficients of the correction formula Eq.~\ref{eq:ml} would have been smaller in compensation.  The greater problem is the loss of halo counts at the low mass end where unseen light either at the extremities of cataloged galaxies or galaxies that are entirely missed impacts the mass function.

It should be possible to evaluate these concerns by "observing" mock catalogs.  There are plans to carry out this experiment.  However ultimately the 2MRS 11.75 catalog is not optimal for a study of the faint end of the mass function because of the loss of low surface brightness systems.  As was discussed in the companion paper T15, it is apparent that there is an abundance of groups with low mass that are entirely missed in the 2MRS 11.75 compilation and would continue to be missed with 2MASS samples to fainter limits.  Surveys that target low surface brightness systems can help clarify the situation at low masses \citep{1959PDDO....2..147V, 1981ApJS...47..139F, 1998A&AS..127..409K, 1999A&AS..135..221K, 2000A&AS..146..359K}.  Wide field neutral Hydrogen surveys provide complementary inventories \citep{2004MNRAS.350.1195M, 2005AJ....130.2598G}.  An anticipated future paper in this series will present a group catalog for galaxies within 3500~\kms\ that samples low surface brightness systems.  That catalog of nearby groups will give much better definition of the low mass end of the mass function.

There can readily be an improvement of the observed mass function at the high mass end as well.  The 2MASS Extended Source Catalog \citep{2000AJ....119.2498J} augmented by Tom Jarrett (private communication) with velocities from the 6dF Galaxy Survey \citep{2014MNRAS.445.2677S} and the Sloan Digital Sky Survey extends a factor 2 greater in depth over 3/4 of the sky, giving access to a factor 6 more volume.

While group catalog extensions to the very local volume and to larger distances will greatly clarify the observational status of the group mass function, there remains an issue with the normalization of the Sheth$-$Tormen mass function.  There has to be a clearer understanding of whether masses in simulations and the observed universe are commensurate.   This issue will be given attention in a future publication.

\section{Summary}

John Huchra's legacy, the 2MASS Redshift Survey with almost all-sky completeness and photometric integrity, provides the best available description of the redshift distribution of galaxies in our corner of the universe \citep{2012ApJS..199...26H, 2006MNRAS.373...45E, 2014MNRAS.445..988N}.  The near-infrared photometry captures light from the dominant baryonic component of galaxies, the old stars, with minimal loss from obscuration.  Variations of this sample have already been used to build group catalogs \citep{2007ApJ...655..790C, 2011MNRAS.416.2840L}.  In the current paper, in addition to using the full 2MRS 11.75 catalog, the groups are constructed to conform to scaling laws defined by detailed studies of individual groups with masses $10^{11}$ to $10^{15}~\Msun$, discussed in T15.

While groups have been defined over the full velocity range of the 2MRS 11.75 catalog, the nearest and furthest groupings are suspect.  Due to the loss of low surface brightness galaxies from the 2MRS catalog, this sample is not optimal for the construction of low mass groups, best studied nearby.  At high redshifts, once only galaxies brighter than the characteristic magnitude $M_K^{\star}$ of the exponential cutoff are being accessed, the correction factor for missing light creates unacceptable uncertainties.  The most useful domain of the present catalog is $3,000-10,000$~\kms\ in the CMB frame.

In this shell from 3,000 to 10,000 \kms, the present catalog contains 24,044 galaxies in 3,461 groups of two or more and 10,144 singles.  The Perseus Cluster, Abell~426, is the most populated group with 180 assigned members.  There are 5 groups with masses greater than $10^{15}~\Msun$ (A2199, Coma, Ophiuchus, Norma, and Perseus), 182 groups in the next decade of mass down to $10^{14}~\Msun$, and 3514 more groups in the subsequent decade down to $10^{13}~\Msun$. The densest concentration of clusters are in the region historically called the "Great Attractor" \citep{1987ApJ...313L..37D}, which is the core of the Laniakea Supercluster \citep{2014Natur.513...71T}, and the Perseus$-$Pisces chain \citep{1988lsmu.book...31H}.

The construction of the group catalog all the way down to "groups" of one permits the construction of the mass function of groups.  The volume considered is between 3,000 and 10,000 \kms\ in the CMB frame, absent 9\% of the sky in the zone of obscuration.  Every 2MRS 11.75 galaxy in this volume contributes to halos as small as $2h^{-1} \times 10^{11}~\Msun$ and as massive as $2h^{-1} \times 10^{15}~\Msun$.  The halo fitting function determined by \citet{1999MNRAS.308..119S} provides a good fit to the shape of the observed mass function with the choice $\Omega_{matter}\sim 0.2$ in a flat universe.  However, the count normalization is poor.  This situation requires further analysis but first there are available ways that the observed mass function can be improved with catalog extensions to the local volume and to greater distances.


The mass in halos in the $3,000-10,000$~\kms\ shell can be summed.  Divided by the volume, the product is a density that can be compared with the critical density for a closed universe without vacuum energy.  This normalized density in bound halos is $\Omega_{collapsed} = 0.16 \pm 0.02$.  The error is the quadrature sums of three components: cosmic variance gauged by the north$-$south difference of $\pm0.009$, 20\% uncertainty in the amplitude of the lost light correction factor resulting in a summed fractional uncertainty of 0.014, and a 10\% uncertainty in the conversion from light to mass, an uncertainty in $\Omega_{collapsed}$ of 0.016..

\bigskip\noindent
{\it Acknowledgements.}  Help in the fitting of mass function formalisms by Julien Carron and Melody Wolk is greatly appreciated.  Daniel Pomar\`ede generated Figs. \ref{2mrsdens_Z}$-$\ref{2mrsdens_X}.   Nick Kaiser and Istvan Szapudi provided useful comments.  This research has been indirectly supported by grants from the US National Science Foundation and the NASA Astrophysics Data Analysis Program.

\bibliography{paper}
\bibliographystyle{apj}

\
\end{document}